# Quantum Communication: From Fundamentals to Recent Trends, Challenges and Open Problems

Hrishikesh Dutta and Amit Kumar Bhuyan
Electrical and Computer Engineering,
Michigan State University, East Lansing, MI, USA

*Abstract* — With the recent advancements and developments in quantum technologies, the emerging field of quantum communication and networking has gained the attention of the researchers. Owing to the unique properties of quantum mechanics, viz., quantum superposition and entanglement, this new area of quantum communication has shown potential to replace modern-day communication technologies. The enhanced security and high information sharing ability using principles of quantum mechanics has encouraged networking engineers and physicists to develop this technology for next generation wireless systems. However, a conceptual bridge between the fundamentals of quantum mechanics, photonics and the deployability of a quantum communication infrastructure is not well founded in the current literature. This paper aims to fill this gap by merging the theoretical concepts from quantum physics to the engineering and computing perspectives of quantum technology. This paper builds the fundamental concepts required for understanding quantum communication, reviews the key concepts and demonstrates how these concepts can be leveraged for accomplishing successful communication. The paper delves into implementation advancements for executing quantum communication protocols, explaining how hardware implementation enables the achievement of all basic quantum computing operations. Finally, the paper provides a comprehensive and critical review of the state-of-the-art advancements in the field of quantum communication and quantum internet; and points out the recent trends, challenges and open problems for the real-world realization of next generation networking systems.

## I. INTRODUCTION

With the recent promising growth of the field of Quantum Computing, there is a high demand for development of technologies that can build networks of these quantum computers. Quantum Communication is a technology that allows interconnection of multiple quantum computers and devices enabling the researchers to build a quantum internet [1] [3]. In addition to obtaining the benefits from the inherent characteristics and properties of quantum computing, such as fast computation speed and parallel processing, quantum communication also offers advantages over classical communication in several aspects. In this early stage of development of this technology, quantum communication has shown promise in ensuring reliable and fast communication with a higher-level security in information sharing and improving the information sharing capacity.

The basic fundamentals of quantum computing and communication lies in quantum mechanics. Quantum mechanics develop physical laws, principles and theories to explain phenomena that cannot be explained by classical laws of physics. These laws and principles with their own unique approaches and explanations about these physical phenomena, have helped uncover, analyze and understand many systems and their properties. Now, building on the theories and concepts of quantum mechanics, this domain of quantum computing and communication has proved to be a useful tool for fast and efficient information processing.

Present-day computing technology has started to reach the quantum-classical boundary. What it means is that the number of chips and transistors are growing day by day and the size of these chips are remarkably reducing following Moore's law [1]. Today we have about 50 billion transistors and the size has reduced to the order of 2 nanometers. Thus, with such miniature devices, it is time to think how the principles of quantum mechanics, that is the theory of micro-world, would affect the behavior of these miniaturized chips and transistors. In short, there has to be new approaches and techniques for processing and communicating information that would meet the needs of this micro-world. While the processing of such information is handled by the technique of quantum computing, as mentioned earlier, the protocols and logics for sharing and communicating information also need revision accordingly. This new domain of communication technology is called quantum communication, with its own new set of rules and principles built from the conceptual basis of quantum mechanics.

Before understanding the working, mechanisms and formal mathematics of this emerging field of quantum communication, the logical question that arises is what are the real benefits that we obtain from this technology. One of the most significant benefits obtained from quantum communication is the higher level of security that is achieved. This enhanced security comes from the fact that the basic unit for quantum information sharing, which we call qubits, possess interesting and unique properties of entanglement and interdependence of states. This allows the communicating entities to detect the presence of any eavesdropper in the network and enables secure information sharing. Researchers have demonstrated that the unique properties of quantum communication significantly enhance communication security over classical communication. Building on this, several key distribution protocols and cryptography techniques have been developed. Another important characteristic of quantum communication is that it allows a higher information sharing capacity as compared to classical communication. This is achieved by the property of quantum systems to encode information with multiple degrees of freedom. In addition to these fundamental advantages, quantum communication also serve many other useful purposes, such as, quantum clock synchronization, distributed quantum information processing and computing, quantum resource sharing etc., which will be explained in later part of this article.

Having said the relevance and significance of this technology, the most logical thought that follows is its practicability and real-world implementation challenges. Although this communication technology is relatively new, researchers around the world have already demonstrated the theoretical findings in hardware setups. To exemplify, the authors in [2] have experimentally validated the point-to-point space to ground quantum communication over a distance of 4600 kilometers. The work in [4] provided a laboratory demonstration of quantum communication that enables communication at a rate that surpasses the ideal loss equivalent direct transmission method. Similarly, the experimental setup using beam-splitter as developed by the authors of [5] show promising results for developing quantum routers. In short, although there are several scopes of improvement in the practical implementation of this technology, the recent trends and developments promise the real-world implementation of quantum communication systems. This has led people to think of the concept of implementation of quantum internet with higher layers of security and fast data processing speed.

As pointed out in the preceding text, quantum communication brings out several advantages over classical communication systems. However, there has to be ways and protocols that allow coexistence of these two communication technologies. And fortunately, there are works [6], [7], as we will see in the later part of this paper, that discuss methods of integrating the two systems. In other words, we will have two internets in the future: classical and quantum internet and they will coexist. To be noted that, as we will see later, quantum communication technology relies on classical communication for sharing the bit strings in the message.

With the main objective of providing an overview and recent trends in the field of quantum communication, this work has three primary contributions. First, this paper provides a detailed tutorial clearly explaining the fundamentals of quantum communication. Starting from the basics of quantum computation, this article will provide an in-depth study of quantum communication. The second contribution of this paper is to provide a detailed and extensive critical literature review of this field, including introductory papers and those with recent ground-breaking works. The third contribution of this work is to point out the open research problems in quantum communication technology. Thus, this paper will serve the purpose of developing a better understanding of the topic for the new researchers in this field, and at the same time, helping them to be aware of the state-of-the-art works in this field, as well as, to find out the open research challenges in this fast-growing technology.

The organization of the paper is as follows. We start with an introduction to quantum communication in Section II, pointing out the distinctions and advantages compared to classical communication. Next, a detailed discussion of quantum computation system is provided in Section III. There we talk about the main components and operations of quantum computing that would help us understand the working of a quantum communication system. This is followed by discussions of the main quantum operations, citing the similarities and differences with their classical counterparts, and quantum measurements. In Section IV, we explain the effect of noise in real world quantum information processing. Section V explains entanglement which is the unique and the most fundamental resource of quantum communication. The next section deal with the practical realization and generation of the quantum states. In Section VII, we explain how point-to-point communication can be established using the concepts of quantum computing technique. This section also provides a detailed description of the different quantum communication protocols proposed in the literature and how these protocols leverage the properties of quantum concepts to achieve the advantages mentioned above. The extension of point-to-point quantum communication to long distance communication is discussed in Section VIII. Section IX talks about the quantum internet, its protocols in different layers and all the current progress from the quantum networking aspects. The next section (Section X) points out the open problems, research challenges and simulation tools developed for implementation of this new technology.

## II. Quantum Communication in the Classical World

Before delving deep into the fundamentals and concepts of quantum communication, we first review the key working a classical communication system. Figure 1 shows the basic building block of a communication system comprising of a transmitter, a channel or a medium for carrying the transmitted message and a receiver. The transmitter encodes the message in suitable form that can be reliably transmitted through the medium and can be decoded by the receiver. The question here is what the best way is to deliver the information or message that the sender wants to convey to the receiver. There are two ways to deliver the message, viz., using analog or digital signals. In analog signal, the sender encodes the message (that can be time-varying) into an analog signal and transmits it directly to the receiver, which, upon receiving the signal, decodes it in order to obtain the information. Digital communication can be executed by sampling these continuous signals and quantizing it into certain discrete intervals and sending these quantized messages. To send these quantized values over the analog channel, the quantized values are encoded to binary strings. These binary strings are composed of '0's and '1's are then converted into analog pulses to be transmitted over the channel. The receiver now receives the pulses, decodes the signal, and estimates the message that was sent by the sender.

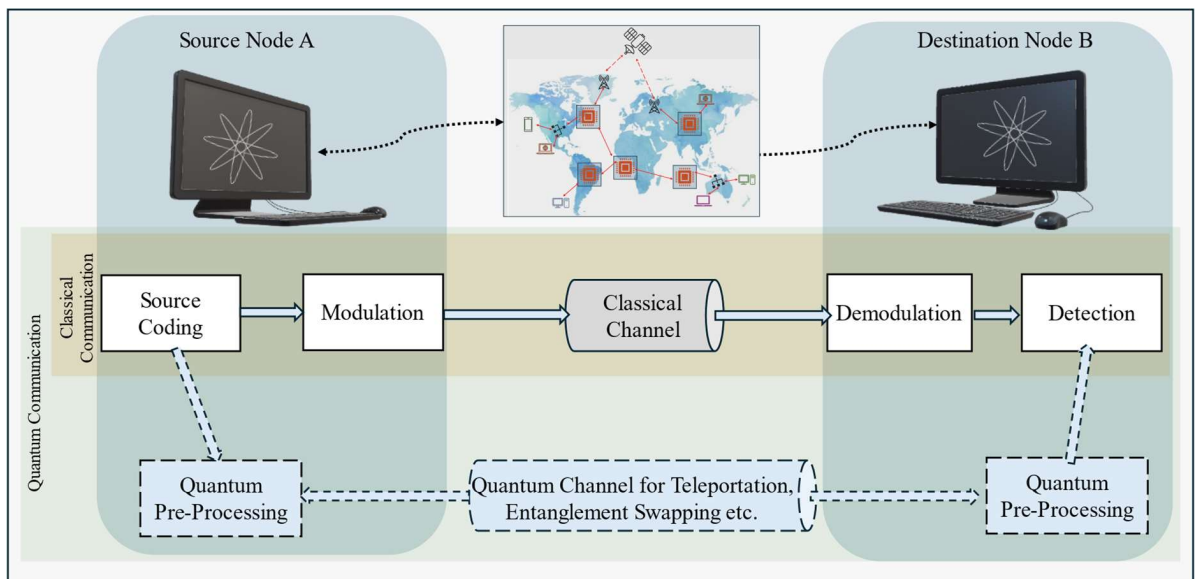

Fig. 1: A high-level representation of Quantum Communication in the classical world

As is evident from the preceding text, the basic unit of information carrier in classical communication is bits. Now, the question is "*Are these bits the most fundamental and only carriers of information?*" In classical communication system, the above statement is true. However, when we enter a quantum communication system, these bits are no longer the basic unit of information carrier. The information carrier in a quantum communication system is called qubit. These qubits have several interesting properties that will be discussed in the following sections, that prove useful to achieve performance better than the classical system.

Now in order to establish communication using qubits, simply having a classical channel does not suffice. As seen in Fig. 1, there is the requirement of a separate channel for accomplishing quantum communication. Quantum channels transmit information using quantum bits or qubits, which can exist in multiple states simultaneously due to principles like superposition and entanglement. Quantum channels can be implemented using various physical systems, including photons, atoms, and superconducting circuits, to carry quantum information over short or long distances. More notably, quantum channels have the potential for high-speed transmission, especially in scenarios involving quantum entanglement and superposition, which enable instantaneous communication between entangled particles regardless of distance.

Nevertheless, we still need classical channels alongside quantum channels for successful implementation of quantum communication systems, due to following reasons. First, classical channels are crucial for controlling and synchronizing quantum operations. They enable the transmission of classical control signals, commands, and timing information that are necessary for initializing quantum states, performing quantum operations, and coordinating quantum processes. Second, classical channels are used to establish the initial parameters and protocols required for quantum communication. For example, classical channels are used to set up the quantum key distribution (QKD) protocol, a popular secure quantum communication protocol. Third, quantum key distribution protocols, such as BB84 and E91, require classical communication channels to exchange initial authentication information and perform key reconciliation and error correction procedures. While the quantum channel is used to transmit the quantum key itself, the classical channel is essential for managing the key distribution process. Quantum error correction codes are typically designed and implemented using classical computing techniques. The results of error detection and correction operations performed on quantum data are communicated back and forth between quantum and classical systems through classical channels. Additionally, after quantum data is transmitted and received, classical channels are used for processing, analyzing, and interpreting the quantum information.

Integrating quantum communication systems with existing classical communication infrastructure requires seamless interaction between classical and quantum channels. Classical channels facilitate the integration and interoperability of quantum technologies with conventional communication networks. In essence, classical channels play a vital role in supporting and complementing quantum channels, enabling the efficient and effective implementation of quantum communication protocols, error correction techniques, and overall system operation. In the next part of this paper, we explain all the fundamentals of quantum communication and protocols. We first start with providing details on a quantum computing and information processing system which would be useful in developing the concepts needed for understanding quantum communication.

### III. AN OVERVIEW OF QUANTUM COMPUTING

#### A. Qubits and Quantum States

The basic building block of quantum computing [8], and hence quantum communication, is a quantum bit, also known as, qubit, denoted by $|\Psi\rangle$. The main distinction of quantum bits from classical bits is that classical bits only have two states, viz., 0 and 1. On the other hand, quantum bits can be $|0\rangle$ or $|1\rangle$ or anything in between. This is represented by the equation:

$$|\Psi\rangle = \alpha|0\rangle + \beta|1\rangle \qquad (1)$$

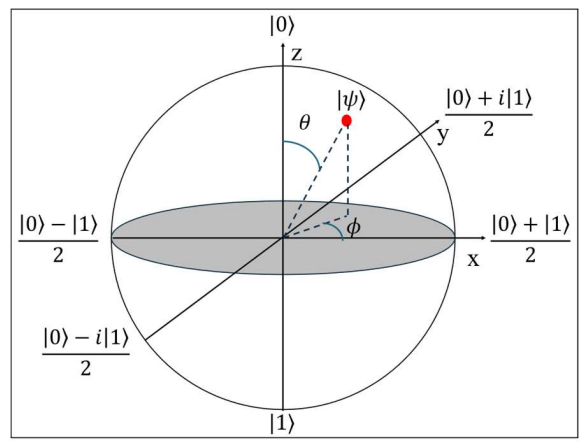

Fig. 2: Representation of a quantum state in Bloch Sphere

Here coefficients $\alpha$ and $\beta$ are complex numbers, known as probability amplitudes, and they are used to determine the probability of the state $|\Psi\rangle$ being in state $|0\rangle$ and $|1\rangle$ respectively. These coefficients should satisfy the normalization condition $\alpha^2 + \beta^2 = 1$. Thus, $\alpha^2$ and $\beta^2$ is the probability of the state $|\Psi\rangle$ being in state $|0\rangle$ and $|1\rangle$ respectively.

Thus, from the definition, we can see that depending on the different values of $\alpha$ and $\beta$, the qubit can have infinitely many states. Two specific scenarios when $\alpha = 0$ and $\beta = 1$ makes the quantum state equal to $|1\rangle$ and $|0\rangle$ respectively, which are known as the basis states [9].

This unique property of qubits, which is, their existence in combination of multiple states simultaneously, is called superposition. As we will see later, this superposition plays a significant role in quantum computing, and particularly, in quantum communication.

A useful representation of a quantum state is the Bloch Sphere [10] [11], as shown in Fig. 2. Bloch Sphere is a visual demonstration of any quantum state, and it is a 3-dimensional sphere. Now any state $|\Psi\rangle$ can be represented by a point on the surface of this sphere. The line connecting this point to the origin has an angle $\theta$ and $\phi$ with two of the axes in the 3-dimensional space. Here, the state $|\Psi\rangle$ can be expressed as

$$|\Psi\rangle = \cos\frac{\theta}{2}|0\rangle + e^{i\phi}\sin\frac{\theta}{2}|1\rangle \quad (2)$$

Here, probability amplitudes of state $|0\rangle$ and $|1\rangle$ are given by $\cos\frac{\theta}{2}$ and $e^{i\phi}\sin\frac{\theta}{2}$ respectively. Now, putting $\theta = 0$ and $\pi$ in the equation we get $|\Psi\rangle = |0\rangle$ and $|1\rangle$ respectively. Thus, as shown in Fig. 2, these states $|0\rangle$ and $|1\rangle$ are located in the north and south pole in the $z$-axis of Bloch Sphere respectively. We can also have an equal superposition of these two basis states, referred to as $|+\rangle$ and $|-\rangle$, defined by:

$$|\pm\rangle = \frac{1}{\sqrt{2}}(|0\rangle \pm |1\rangle)$$

These two states lie on the two extremes of another orthogonal axis of the Bloch Sphere ($\theta = \frac{\pi}{2}, \phi = 0$). Similarly, the two states that are located on the two extremes of the sphere are denoted as $|+i\rangle$ and $|-i\rangle$ and are also an equal superposition of $|0\rangle$ and $|1\rangle$. For these two superposed states, the phase between $|0\rangle$ and $|1\rangle$ is $\phi = \frac{\pi}{2}$ and can be denoted as follows.

$$|\pm i\rangle = \frac{1}{\sqrt{2}}(|0\rangle \pm i|1\rangle)$$

Note that the equator of the sphere represents the states with equal probabilities of being in state $|0\rangle$ and $|1\rangle$.

### B. Quantum Operations

The next fundamental point that needs attention is the mechanism to handle these qubits. In other words, the question that we need to answer is how to process these quantum states in order to accomplish meaningful transformations and operations. The following are some basic quantum transformations [12] [13] to start with.

*Identity Operation:* This is the most basic operation, which essentially keeps the qubits as they are. This is similar to the identity transformation in the classical world that operates on classical bits 0 and 1, without changing their states. This operation is represented as:

$$|\Psi\rangle \xrightarrow{I} |\Psi\rangle$$

We represent the identity operator in matrix form as $I = \begin{bmatrix} 1 & 0 \\ 0 & 1 \end{bmatrix}$

*Flip Operation:* This is analogous to the NOT operator in classical world. It transforms $|0\rangle$ to $|1\rangle$ and $|1\rangle$ to $|0\rangle$. The operator is called Pauli-$X$ operator and is given by:

$$X = \begin{bmatrix} 0 & 1 \\ 1 & 0 \end{bmatrix}$$

$$|0\rangle \xrightarrow{X} |1\rangle \text{ and } 1\rangle \xrightarrow{X} |0\rangle$$

*Hadamard Operation:* This is one of the most useful operations in quantum mechanics. It operates on qubits $|0\rangle$ and $|1\rangle$, and produces an equal superposition of these states [14] [15]. This operation [16] does not have a classical counterpart and it is represented formally as:

$$H = \frac{1}{\sqrt{2}}\begin{bmatrix} 1 & 1 \\ 1 & -1 \end{bmatrix}$$

$$|0\rangle \xrightarrow{H} \frac{|0\rangle+|1\rangle}{\sqrt{2}} \text{ and } |1\rangle \xrightarrow{H} \frac{|0\rangle-|1\rangle}{\sqrt{2}}$$

*Rotation Operation:* This operator, parameterized by $\hat{n}$ and $\theta$, transforms any quantum state $\Psi$ by rotating by angle $\theta$ in a plane defined by vector $\hat{n}$. This is formally defined as:

$$R_{\hat{n}}(\theta) = e^{-\theta.\hat{n}.\frac{\hat{\sigma}}{2}}$$

As for example, rotating the state $|0\rangle$ by angle $\frac{\pi}{2}$ along axis $Y$ gives the state $|+\rangle$ (Fig. 2). That is,

$$R_Y\left(\frac{\pi}{2}\right)|0\rangle = |+\rangle$$

Two other commonly used quantum operators are Pauli-$Y$ and Pauli-$Z$ operators defined as:

$$Y = \begin{bmatrix} 0 & -i \\ i & 0 \end{bmatrix} \text{ and } Z = \begin{bmatrix} 1 & 0 \\ 0 & -1 \end{bmatrix}$$

These two operators physically mean flipping of the quantum state around Pauli-Y and Pauli-Z axes in the Bloch sphere respectively. Note that, these operators are related to the rotation operator as represented as:

$$-iY = R_Y(\pi) \text{ and } -iZ = R_Z(\pi)$$

Note that all these operators that we have seen above bears the following common property. Any of the above operations $U$ can be reversed using its adjoint $U^\dagger$. That is, if a quantum state $|\Psi\rangle$ is transformed using these operators to get $|\Psi'\rangle$, we can get the original state simply by operating $|\Psi'\rangle$ with the adjoint of that operator. Mathematically,

$$|\Psi\rangle = U^\dagger|\Psi'\rangle = U^\dagger U|\Psi\rangle \quad (3\ a)$$

Also note that,

$$UU^\dagger = I \quad (3\ b)$$

There is a specific name for those operators that satisfy Eqns. (3 a) and (3 b) and they are called Unitary Operators.

There is another class of operators frequently used in quantum computing, known as Hermitian Operators [17]. The characteristic property of such operators are that the matrices representing these operators are adjoint to themselves. Mathematically for a Hermitian Operator $H$,

$$H = H^\dagger \quad (4)$$

Examples of Hermitian Operators are (a) Projection Operators ($P = |\psi\rangle\langle\psi|$), (b) Hamiltonian Operator ($\hat{H}$) that represents the total energy in a quantum system ($i\hbar\frac{\partial\psi}{\partial t} = \hat{H}\psi$), to name a few. While here we stop by just defining the Hermitian operator, we will provide the details of any Hermitian operation whenever we require to explain any phenomena requiring the usage of this operation later in the text.

As will be shown later, these Unitary and Hermitian transformations play a vital role in quantum communications, including enabling quantum operations and measurements and also in setting up of the most unique quantum entanglement property.

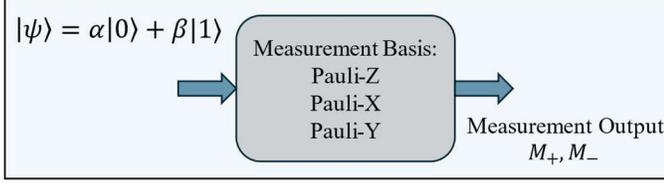

Fig. 3: Measurement of quantum states

### C. Quantum Measurements

Till now, we have understood what a quantum state is and how to process it using different unitary operations. Let's say, we are given a qubit $\psi = \alpha|0\rangle + \beta|1\rangle$ after certain operations performed on it. Then we need some ways to extract information out of the qubit. In other words, we need to address how to know in what state the qubit was prepared. For example, one might want to know if it was prepared in state $|0\rangle$ or $|1\rangle$. This process of inference from a qubit is called measurement. Measurement [18] [19] can be seen as a black box where the input is a qubit, and the output is the state of the qubit (Fig. 3). If we code the output measurement variables as $M_+$ and $M_-$ respectively where one of them indicate that the qubit being in state $|0\rangle$ and the other indicate that it is in state $|1\rangle$ respectively, then the probability of getting each outcome can be expressed this as:

$P[M_+] = |\alpha|^2$: The qubit is in state $|0\rangle$
$P[M_-] = |\beta|^2$: The qubit is in state $|1\rangle$

To be noted here that once the measurement of the qubit is done, its state changes. In the above example, if the measurement outcome is $M_+$, we know for sure that the qubit is in state $|0\rangle$. This means that the state of the qubit has changes from $|\psi\rangle$ to $|0\rangle$. And the probability of the occurrence of this outcome is $P[M_+] = |\alpha|^2$. Similarly, if the measurement outcome is $M_-$, we know for sure that the state of the qubit is $|1\rangle$ and thus the state changes to $|1\rangle$ post measurement. This happens with probability $|\beta|^2$. This particular measurement in the above example, where we are looking for $|0\rangle$ and $|1\rangle$ states, is a measurement performed in the Pauli-Z basis. This is because the states $|0\rangle$ and $|1\rangle$ lie in the Pauli-Z axis (defined by $|0\rangle$ and $|1\rangle$) in the Bloch sphere mentioned above.

To generalize the above concept, we can state it formally as follows. When a measurement for a state $\psi = \alpha|0\rangle + \beta|1\rangle$ is done in a basis $\phi$ defined by $<\phi_0, \phi_1>$, then, the probability of occurrence of $M_+$ is given by the squared magnitude of inner product of state $\psi$ with $\phi_0$ and $\phi_1$ respectively. That is,

$P[M_+] = |\langle\phi_0|\psi\rangle|^2$ and $P[M_-] = |\langle\phi_1|\psi\rangle|^2$ (5)

Thus, in the above example, where the measurement basis is Pauli-Z (defined by $|0\rangle$ and $|1\rangle$), we get the probabilities as:

$P[M_+]_Z = |\langle\phi_0|\psi\rangle|^2 = |\langle 0|\psi\rangle|^2$
$= |(1 \quad 0)\begin{pmatrix}\alpha\\\beta\end{pmatrix}|^2$
$= |\alpha|^2$

Similarly, we can obtain $P[M_-]_Z = |\beta|^2$.

Along the same lines, measurements can be done in Pauli-X and Pauli-Y basis respectively, where the probabilities of getting $M_+$ and $M_-$ are given by:

Pauli-X: $P[M_\pm]_X = \frac{|\alpha \pm \beta|^2}{2}$

Pauli-Y: $P[M_\pm]_Y = \frac{|\alpha \mp i\beta|^2}{2}$

One important point to add here would be to give a formal definition of a basis that we use to perform quantum measurement. Basis is a set of orthonormal vectors that can be used to express any other vector. To exemplify, the following are the basis vectors that we have used in the explanation above.

Z-Basis: $\{|0\rangle = \begin{pmatrix}1\\0\end{pmatrix}; |1\rangle = \begin{pmatrix}0\\1\end{pmatrix}\}$

X-Basis: $\{|+\rangle = \frac{1}{\sqrt{2}}\begin{pmatrix}1\\1\end{pmatrix}; |-\rangle = \frac{1}{\sqrt{2}}\begin{pmatrix}1\\-1\end{pmatrix}\}$

Y-Basis: $\{|+i\rangle = \frac{1}{\sqrt{2}}\begin{pmatrix}1\\-i\end{pmatrix}; |-i\rangle = \frac{1}{\sqrt{2}}\begin{pmatrix}1\\-i\end{pmatrix}\}$

The above explanation of measurement of quantum state $|\psi\rangle$ in a basis $\phi$ defined by $<\phi_0, \phi_1>$ can also be represented by the outer-product operation as follows:

$\Pi_+^\phi|\psi\rangle = |\phi_0\rangle\langle\phi_0||\psi\rangle$ and $\Pi_-^\phi\psi = |\phi_1\rangle\langle\phi_1||\psi\rangle$ (6)

If we consider the above example of measuring $|\psi\rangle$ is Pauli-Z (defined by $|0\rangle$ and $|1\rangle$), we can obtain the measurement outcome from Eqn. (6) as:

$\Pi_+^Z|\psi\rangle = |0\rangle\langle 0|(\alpha|0\rangle + \beta|1\rangle) = \alpha|0\rangle\langle 0|0\rangle + \beta|0\rangle\langle 0|1\rangle$
$= \alpha|0\rangle$

Similarly,
$\Pi_-^Z|\psi\rangle = \beta|1\rangle$

This tells us that measurement of quantum state $|\psi\rangle$ in Pauli Z basis projects it to $|0\rangle$ and $|1\rangle$ with probabilities $|\alpha|^2$ and $|\beta|^2$ respectively, which are the same measurement probabilities in Pauli-Z basis for obtaining $M_+$ and $M_-$ as the measurement outcomes. We observed the same results by using Eqn. (5) for measurement.

Now that we know how to compute the probability of getting a particular measurement outcome from a given state and a given basis $\phi$, we can find out all sorts of statistical measurements from it. For example, if the measurement is done in $X$-basis, one might be interested in knowing the expectation of the measurement outcome. This can be computed as follows, using basic probability theory:

$\mathbb{E}[X] = P[M_+]_X \times M_+ + P[M_-]_X \times M_-$
$= \frac{|(\alpha+\beta)|^2}{2} \times M_+ + \frac{|(\alpha-\beta)|^2}{2} \times M_-$

The above expectation can be found for specific values of measurement values. One common scheme of assigning measurement outcome values is assigning $+1$ to $M_+$ and $-1$ to $M_-$. Using this scheme, we can rewrite the expectation as:

$\mathbb{E}[X] = \frac{|(\alpha+\beta)|^2}{2} - \frac{|(\alpha-\beta)|^2}{2}$

Since $\alpha, \beta \in \mathbb{C}$,

$\mathbb{E}[X] = \frac{1}{2}((\alpha+\beta)(\alpha+\beta)^* - (\alpha-\beta)(\alpha-\beta)^*)$

$= \frac{1}{2}(|\alpha|^2 + |\beta|^2 + \alpha^*\beta + \alpha\beta^* - |\alpha|^2 - |\beta|^2 + \alpha^*\beta + \alpha\beta^*)$

$\Rightarrow \mathbb{E}[X] = \alpha^*\beta + \alpha\beta^*$ (7)

The variance in measurement can be computed as:
$V(X) = \mathbb{E}[X^2] - (\mathbb{E}[X])^2$
$= P[M_+]_X \times M_+^2 + P[M_-]_X \times M_-^2$
$\quad - (P[M_+]_X \times M_+ + P[M_-]_X \times M_-)^2$

$$= \frac{|(\alpha+\beta)|^2}{2} \times M_+^2 + \frac{|(\alpha-\beta)|^2}{2} \times M_-^2$$
$$- \left( \frac{|(\alpha+\beta)|^2}{2} \times M_+ + \frac{|(\alpha-\beta)|^2}{2} \times M_- \right)^2$$

Putting $M_+ = +1$ and $M_- = -1$, we get
$$Var(X) = \frac{|(\alpha+\beta)|^2}{2} + \frac{|(\alpha-\beta)|^2}{2}$$
$$- \left( \frac{|(\alpha+\beta)|^2}{2} - \frac{|(\alpha-\beta)|^2}{2} \right)^2$$

Using the above logic, the expectation and variance of measurement outcome in any basis $\phi$ can be computed:
$$\mathbb{E}[\phi] = P[M_+]_\phi \times M_+ + P[M_-]_\phi \times M_-$$
$$Var(\phi) = \mathbb{E}[\phi^2] - (\mathbb{E}[\phi])^2$$

The expectation can be computed using linear algebra as follows:
$$\mathbb{E}[\phi] = \langle \phi \rangle = \langle \psi | \phi | \psi \rangle \qquad (8)$$

This can be easily verified using the above example of measurement in $X$ basis:
$$\mathbb{E}[X] = \langle X \rangle = \langle \psi | X | \psi \rangle$$
$$= (\alpha^* \;\; \beta^*) \begin{bmatrix} 0 & 1 \\ 1 & 0 \end{bmatrix} \begin{pmatrix} \alpha \\ \beta \end{pmatrix}$$
$$= (\alpha^* \;\; \beta^*) \begin{pmatrix} \beta \\ \alpha \end{pmatrix}$$
$$= \alpha^*\beta + \alpha\beta^*$$

The above expression is the same as is obtained in Eqn. (7). In the Dirac notation as is done above, the variance can be expressed as:
$$Var(\phi) = (\Delta\phi)^2$$
$$Var(\phi) = \langle \phi^2 \rangle - (\langle \phi \rangle)^2 \qquad (9)$$

Thus, using equations (8) and (9), the expectation and variance of measurement in any basis $\phi$ can be computed.

### D. Multiple Qubits

Now let us extend the above concept to a scenario of multiple qubits. Let us first bring the analogy with classical bits. Using '$b$' number of *classical bits*, the number of states that can be obtained is $2^b$ which can be represented as follows.

$$0000\ldots00, 0000\ldots01, \ldots, 0000\ldots10, \ldots, 1111\ldots11$$

On the other hand, using '$b$' number of *qubits*, we can obtain states which can be represented in the following form:
$$\psi = \alpha_1 |00\ldots00\rangle + \alpha_2 |00\ldots01\rangle + \ldots + \alpha_{2^b} |11\ldots11\rangle \qquad (10)$$

Here, $|00\ldots00\rangle, |00\ldots01\rangle, \ldots, |11\ldots11\rangle$ are the basis states that can be defined by unit vectors of length $2^b$ as:

$$|00\ldots00\rangle = \begin{pmatrix} 1 \\ 0 \\ 0 \\ \vdots \\ 0 \end{pmatrix}, |00\ldots01\rangle = \begin{pmatrix} 0 \\ 1 \\ 0 \\ \vdots \\ 0 \end{pmatrix}, \ldots, |11\ldots11\rangle = \begin{pmatrix} 0 \\ 0 \\ 0 \\ \vdots \\ 1 \end{pmatrix} \qquad (11)$$

Note that similar to the case with $b = 1$, here also the normalization principle holds:
$$\sum_{i=1}^{2^b} \alpha_i^2 = 1 \qquad (12)$$

The $b$-qubit state denoted in Eqn. (10) can also be obtained if we know the state of each individual qubit. If the state of each individual qubit is given by $|a_i\rangle, i = 1,2,3,\ldots b$, then the state of $\psi$ in Eqn. (10) can be expressed as the tensor product of $|a_i\rangle$'s as below:

State of $|\psi\rangle = |a_1\rangle \otimes |a_2\rangle \otimes |a_3\rangle \otimes \ldots |a_b\rangle \qquad (13)$

As an example, for a 2-qubit system, if one qubit is in state $|1\rangle$ and the other qubit is in $|+\rangle$, then the 2-qubit state of $|\psi\rangle$ can be written as:

$$|1\rangle \otimes |+\rangle = \begin{pmatrix} 0 \\ 1 \end{pmatrix} \otimes \frac{1}{\sqrt{2}} \begin{pmatrix} 1 \\ 1 \end{pmatrix} = \frac{1}{\sqrt{2}} \begin{pmatrix} 0 \\ 0 \\ 1 \\ 1 \end{pmatrix}$$

In this way, all the properties and operations that we have developed for a single-qubit system applies to this system of multiple qubits.

*Operators on Multiple Qubits*: There are quantum operators that act on multiple qubits. One such multi-qubit operator is the CNOT gate. The CNOT gate acts on a two-qubit state $|c_1 c_2\rangle$ and converts it to $|c_1 c_2'\rangle$, where $|c_2'\rangle$ is dictated by the control qubit $|c_1\rangle$. The CNOT operation on a two-qubit state can be given by the following examples:
$$|00\rangle \xrightarrow{CNOT} |00\rangle, |01\rangle \xrightarrow{CNOT} |01\rangle, |10\rangle \xrightarrow{CNOT} |11\rangle \text{ and}$$
$$|11\rangle \xrightarrow{CNOT} |10\rangle$$

Formally, the CNOT operator/gate can be expressed as:
$$CNOT = I \otimes |0\rangle\langle 0| + X \otimes |1\rangle\langle 1| \qquad (13)$$

In matrix form, this can be represented as
$$CNOT \equiv \begin{bmatrix} 1 & 0 & 0 & 0 \\ 0 & 1 & 0 & 0 \\ 0 & 0 & 0 & 1 \\ 0 & 0 & 1 & 0 \end{bmatrix}$$

This operator plays an important role in quantum computing and information processing. CNOT gates can be used for quantum entanglement creation, which is a fundamental resource for establishing secure quantum communication. CNOT has also found its place in quantum error correction and other algorithms which will be covered in this paper.

## IV. QUANTUM COMPUTING IN REAL WORLD: DEALING WITH NOISE

### A. Sources of Noise

Till this point in this article, we have understood what a qubit and a quantum state are; what is the relationship between them; how to process and extract information from those quantum states. While doing so, we have assumed a noise-free scenario. What that means is that till now, we assumed that the quantum state that we have prepared is the exact quantum state that we want it to be. In other words, we have full knowledge about the quantum state that we are dealing with.

However, unfortunately, this is not the case when handling quantum states in the real world. In fact, while preparing a quantum state, we cannot deterministically prepare a quantum state $|\psi\rangle$. Instead, what we prepare a set of states $\{|\psi_1\rangle, |\psi_2\rangle, \ldots, |\psi_N\rangle\}$ with a probability distribution of $\{p_1, p_2, \ldots, p_N\}$. In physical sense, it means that we do not have perfect knowledge about a quantum state even right after its preparation. What we can only know is the probability of $|\psi\rangle$ being in a particular state.

There are other sources of noise as well. One example is the noise arising from the errors in processing in quantum computers. The unitary operation that we intend the processor to perform on a quantum state may not be the exact operation that the processor/computer actually performs. These kinds of errors can be classified into two categories: coherent errors and incoherent errors. In coherent errors the operation performed by the processor on the quantum state is unitary whereas, for the latter, the performed operation is not unitary.

The third category of noise is the most important one from the point of view of quantum communication. This kind of noise arises due to imperfection in the communication channel. These noises can be both coherent and incoherent. There may be other scenarios of amplitude attenuation of the signals carrying the quantum states.

We will see later in this article how reliable quantum communication can be achieved even in the presence of these errors and noise. But what we really want to see immediately after this is how to analyze or extract information from the quantum states in the presence of noise and errors.

### B. Noisy Quantum States: Pure and Mixed States

As we have observed in the previous discussion, it is close to impossible to have the perfect knowledge of a quantum state $|\psi\rangle$ [20]. Instead, what we know is the distribution of the quantum states. This scenario, where we have a mixture of many states $\{|\psi_1\rangle, |\psi_2\rangle, \ldots, |\psi_N\rangle\}$ with a probability distribution of $\{p_1, p_2, \ldots, p_N\}$, can be represented by a density matrix $\rho$ as:

$$\rho = \sum_{i=1}^{N} p_i |\psi_i\rangle\langle\psi_i| \quad (14)$$

To understand, what this physically means, let us consider an example, where we intend to send a quantum state $|\psi\rangle$ over a noisy communication channel that flips the input state with a probability of $p_x$. Now the signal at the receiver will have a mixture of the original state $|\psi\rangle$ and its flipped version, that is $X|\psi\rangle$. Here the output can be represented by a density matrix as follows:

$$\rho = (1 - p_x)|\psi\rangle\langle\psi| + p_x X|\psi\rangle\langle\psi|X$$

When $|\psi\rangle = |0\rangle$, the density matrix becomes:

$$\rho = (1 - p_x)|0\rangle\langle 0| + p_x |1\rangle\langle 1| = \begin{bmatrix} 1 - p_x & 0 \\ 0 & p_x \end{bmatrix}$$

When $|\psi\rangle = |1\rangle$, the density matrix becomes:

$$\rho = (1 - p_x)|1\rangle\langle 1| + p_x |0\rangle\langle 0| = \begin{bmatrix} p_x & 0 \\ 0 & 1 - p_x \end{bmatrix}$$

Such a noisy, non-deterministic quantum state which needs to be expressed by a density matrix $\rho$ and which cannot be expressed as a single state $|\psi\rangle$ or a superposition of multiple states, is called a *Mixed State* [21] [22]. Note that the density matrix has the property that it has to follow the normalization condition, that is, the trace of the density matrix equals to 1.

$$Tr(\rho) = 1 \quad (15)$$

Also, observe that, the density matrix $\rho$ satisfies both Unitary and Hermitian property [23]: $\rho\rho^\dagger = I$ and $\rho = \rho^\dagger$ (See Section *III B*).

The other category of quantum states which can be represented as a single state $|\psi\rangle$ or a superposition of multiple states, and which assumes that the state is known with certainty, is called a *Pure State*.

Note that a *Mixed State* needs to be represented by a matrix instead of a vector which can be used to represent a *Pure State*, as we have seen in the previous example.

Another distinguishing property of these two types of quantum states from the Bloch Sphere perspective is that any pure state $|\psi\rangle$ lies on the surface of the Bloch Sphere. However, any mixed state resides inside the Bloch Sphere. [24] [25] In other words, the distance of the point representing the mixed state from the center of the Bloch sphere is less than 1. For the extreme case, a state defined by $\rho$ that lies on the center of the sphere is called a *Maximally Mixed State*. [26] The interesting point to note here is that this *Maximally Mixed State* is an equal superposition of two pure states that lie in the opposite poles on the surface of the Bloch Sphere. For example, the *Maximally Mixed State* ($\rho_{MM}$) can be represented as:

$$\rho_{MM} = \frac{1}{2}(|0\rangle\langle 0| + |1\rangle\langle 1|) = \frac{1}{2}(|+\rangle\langle +| + |-\rangle\langle -|)$$
$$= \frac{1}{2}(|+i\rangle\langle +i| + |-i\rangle\langle -i|) \quad (16)$$

From the above definition, we can also identify the difference between the *Maximally Mixed State* ($\rho_{MM}$) [27] and a superposed state (which is a pure state!). In other words, measurement of $\rho_{MM}$ in any basis $\phi$ will give the probabilities $P[M_+] = P[M_-] = 0.5$. On the other hand, for superposed states, these probabilities vary with the basis used for measurement. For example, $|+\rangle = \frac{|0\rangle + |1\rangle}{\sqrt{2}}$ is an equal superposition of $|0\rangle$ and $|1\rangle$ and its measurement in Z-basis results in probabilities $P[M_+^Z] = P[M_-^Z] = 0.5$. On the other hand, if we measure in X-basis, we obtain $P[M_+^X]1$ and $P[M_-^X] = 0$.

What it means in short, physically, is that, if we are given a pure state, we have the perfect knowledge of the state, whereas, for mixed states, we do not know for sure what the quantum state is. We can only know (stochastically) the probability that the given mixed state is in a particular state.

### C. Metric for Measuring Mixture of States: Fidelity

As we have seen, *Pure States* are difficult to obtain, what we need is a measure to denote how different the *Mixed State* is from the desired *Pure State*. This is represented by an index called Fidelity [28] [29], which we define mathematically as follows:

$$F(\rho, |\psi\rangle) = \langle\psi|\rho|\psi\rangle \quad (17)$$

Here, $\rho, |\psi\rangle$ are the *Mixed State* and the desired *Pure State* respectively. The values of Fidelity $F$ lies between 0 and 1, both inclusive.

If the *Mixed State* and the desired *Pure State* are the same, then the Fidelity $F(\rho, |\psi\rangle) = 1$. If the *Mixed State* is orthogonal to the desired *Pure State* are the same, then the Fidelity $F(\rho, |\psi\rangle) = 0$. On the other hand, if *Mixed State* is maximally mixed, then the Fidelity $F(\rho, |\psi\rangle) = \frac{1}{2^N}$, where $N$ is the number of qubits used to represent the *Maximally Mixed State* [30].

To give an example, let us consider the flip channel that we used to exemplify the *Mixed State* in Subsection *IV (C)*. We can find the fidelity values for that example using Eqn. (17) as follows:

$$F(\rho, |\psi\rangle) = \langle\psi|\rho|\psi\rangle$$
$$= \langle\psi|((1 - p_x)|\psi\rangle\langle\psi| + p_x X|\psi\rangle\langle\psi|X)|\psi\rangle$$
$$F(\rho, |0\rangle) = \langle 0|\rho|0\rangle$$
$$= (1 - p_x)\langle 0|0\rangle\langle 0|0\rangle + p_x \langle 0|1\rangle\langle 1|0\rangle$$
$$= (1 - p_x).1 + p_x.0 = 1 - p_x$$
$$\text{And } F(\rho, |1\rangle) = \langle 1|\rho|1\rangle$$
$$= (1 - p_x)\langle 1|1\rangle\langle 1|1\rangle + p_x \langle 1|0\rangle\langle 0|1\rangle$$
$$= (1 - p_x).1 + p_x.0 = 1 - p_x$$

This tells us that if the noise of the communication channel is high, that is $p_x \to 1$, we get $F(\rho, |\psi\rangle) \to 0$, meaning the output state is completely orthogonal to the desired *Pure State*. Similarly, if the channel is very reliable, that is $p_x \to 0$, we get $F(\rho, |\psi\rangle) \to 1$, meaning the output *Mixed State* is close to the desired *Pure State*.

## V. ENTANGLEMENT: THE RESOURCE FOR QUANTUM COMMUNICATION

### A. What is Entanglement?

One of the most useful concepts of quantum mechanics that is used in quantum communication is entanglement. [31] [32] [33] This is a unique property of quantum mechanics leveraged for achieving security in quantum communication. Let us understand quantum entanglement using an example of two scenarios.

*Scenario 1:* Let us consider two entities $A$ and $B$ that have local states $|\psi_A\rangle = a_0|0\rangle + a_1|1\rangle$ and $|\psi_B\rangle = b_0|0\rangle + b_1|1\rangle$ respectively. These 'entities' can be anything, ranging from quantum particles to a device that can do quantum computing. Let us consider that the global quantum state of the system comprising of entities $A$ and $B$ is $|\psi_{AB}\rangle = |01\rangle$.

Now using Eq. (13), we can write the global state of the two-qubit system as:
$$|\psi_{AB}\rangle = |\psi_A\rangle \otimes |\psi_B\rangle$$
$$\Rightarrow |01\rangle = (a_0|0\rangle + a_1|1\rangle) \otimes (b_0|0\rangle + b_1|1\rangle)$$
$$\Rightarrow |01\rangle = a_0 b_0|00\rangle + a_0 b_1|01\rangle + a_1 b_0|10\rangle + a_1 b_1|11\rangle$$

By comparing the left- and right-hand sides of the above equation, we get
$$a_0 b_0 = 0, a_0 b_1 = 1, a_1 b_0 = 0, a_1 b_1 = 0 \quad (14)$$

The solution of the sets of equations in (14) is:
$$a_1 = b_0 = 0; a_0 = b_1 = 1$$

From this, we know the local states of entities $A$ and $B$ are $|\psi_A\rangle = |0\rangle$ and $|\psi_B\rangle = |1\rangle$.

Thus, in this scenario, we can deterministically predict the local states of two entities, if we know the global quantum state of the system.

*Scenario 2:* In this case, like the previous scenario, let us consider two entities $A$ and $B$ that have local states $|\psi_A\rangle = a_0|0\rangle + a_1|1\rangle$ and $|\psi_B\rangle = b_0|0\rangle + b_1|1\rangle$ respectively. However, in this case, the global state of the system, let's say, is known to be $|\psi_{AB}\rangle = \frac{1}{\sqrt{2}}(|01\rangle + |10\rangle)$.

Now using the same procedure used in scenario 1, we can write the following equation:
$$|\psi_{AB}\rangle = |\psi_A\rangle \otimes |\psi_B\rangle$$
$$\Rightarrow \frac{1}{\sqrt{2}}(|01\rangle + |10\rangle) = (a_0|0\rangle + a_1|1\rangle) \otimes (b_0|0\rangle + b_1|1\rangle)$$
$$\Rightarrow \frac{1}{\sqrt{2}}(|01\rangle + |10\rangle) = a_0 b_0|00\rangle + a_0 b_1|01\rangle + a_1 b_0|10\rangle + a_1 b_1|11\rangle$$

By comparing the left- and right-hand sides of the above equation, we get
$$a_0 b_0 = 0, a_0 b_1 = \tfrac{1}{\sqrt{2}}, a_1 b_0 = \tfrac{1}{\sqrt{2}}, a_1 b_1 = 0 \quad (18)$$

Observe that, solutions to the set of equations given by (18) do not exist. In other words, given a global state of the system defined by $|\psi_{AB}\rangle = \frac{1}{\sqrt{2}}(|01\rangle + |10\rangle)$, we cannot find out the local state of each individual entity $|\psi_A\rangle$ and $|\psi_B\rangle$.

Such a global state (the one in *Scenario 2*) is called as an entangled state, and this phenomenon is called entanglement.

Now we are in a position to formally define *Entanglement*. Any global state that cannot be written as the tensor-product of individual local states is known as *Entangled State*. Mathematically, an *Entangled State* can be written as:
$$|\psi_{AB}\rangle \neq |\psi_A\rangle \otimes |\psi_B\rangle \quad (19)$$

The other category of '*non-Entangled State*' (*scenario 1*) is called *Product state* [34]. In a more formal way, *Product state* is one that represents that perfect knowledge of the global state denotes perfect knowledge of the individual local states. Mathematically, a *Product State* can be written as:
$$|\psi_{AB}\rangle = |\psi_A\rangle \otimes |\psi_B\rangle \quad (20)$$

To understand the benefit of entanglement at an abstract level, let us consider in the example of *Scenario 2* that entities $A$ and $B$ are nodes that are communicating with each other. If these nodes share an entangled state $|\psi\rangle$ with each other, then using that global entangled state, an eavesdropper cannot infer the local states of nodes $A$ and $B$. This is one of the key fundamental theories that is used for enhancing security in quantum communication over classical communication. All the details and explanation about achieving such secure quantum communication using shared entangled states will be discussed while we cover the protocols for quantum communication later in this paper.

### B. Bell Basis

The class of entangled quantum states with two qubits is called Bell States [35] [36] and has a special role to play in quantum communication. There are four Bell states:
$$|\phi^+\rangle = \frac{1}{\sqrt{2}}(|00\rangle + |11\rangle)$$
$$|\phi^-\rangle = \frac{1}{\sqrt{2}}(|00\rangle - |11\rangle)$$
$$|\psi^+\rangle = \frac{1}{\sqrt{2}}(|01\rangle + |10\rangle)$$
$$|\psi^-\rangle = \frac{1}{\sqrt{2}}(|01\rangle - |10\rangle)$$

An interesting property about these Bell States is that they form an orthogonal basis of two qubits [37] [38]. Thus, the computational basis for two qubits can be expressed in terms of these Bell States as follows:
$$|00\rangle = \frac{1}{\sqrt{2}}(|\phi^+\rangle + |\phi^-\rangle)$$
$$|11\rangle = \frac{1}{\sqrt{2}}(|\phi^+\rangle - |\phi^-\rangle)$$
$$|01\rangle = \frac{1}{\sqrt{2}}(|\psi^+\rangle + |\psi^-\rangle)$$
$$|10\rangle = \frac{1}{\sqrt{2}}(|\psi^+\rangle - |\psi^-\rangle)$$

Now that we have another orthogonal computation basis, all the different kinds of operations that we studied in Section IV can be extended for Bell basis as well. Any two-qubit state $|\psi\rangle$ can be represented in Bell basis as:
$$|\psi\rangle = \alpha_{00}|00\rangle + \alpha_{01}|01\rangle + \alpha_{10}|10\rangle + \alpha_{11}|11\rangle$$
$$\Rightarrow |\psi\rangle = \frac{\alpha_{00}}{\sqrt{2}}(|\phi^+\rangle + |\phi^-\rangle) + \frac{\alpha_{01}}{\sqrt{2}}(|\psi^+\rangle + |\psi^-\rangle)$$
$$+ \frac{\alpha_{10}}{\sqrt{2}}(|\psi^+\rangle - |\psi^-\rangle) + \frac{\alpha_{11}}{\sqrt{2}}(|\phi^+\rangle - |\phi^-\rangle)$$

$$\Rightarrow |\psi\rangle = \frac{\alpha_{00}+\alpha_{11}}{\sqrt{2}}|\phi^+\rangle + \frac{\alpha_{01}+\alpha_{10}}{\sqrt{2}}|\psi^+\rangle + \frac{\alpha_{00}-\alpha_{11}}{\sqrt{2}}|\phi^-\rangle + \frac{\alpha_{01}-\alpha_{10}}{\sqrt{2}}|\psi^-\rangle \quad (21)$$

Using Eqn. (21), any two-qubit quantum state can be expressed in Bell State, and this indicates that we can perform the measurement in Bell basis, similar to the measurements done in Pauli-Z, X or Y basis. The measurement probabilities in this case becomes:

$$P[|\phi^+\rangle] = \left|\frac{\alpha_{00}+\alpha_{11}}{\sqrt{2}}\right|^2$$

$$P[|\phi^-\rangle] = \left|\frac{\alpha_{00}-\alpha_{11}}{\sqrt{2}}\right|^2$$

$$P[|\psi^+\rangle] = \left|\frac{\alpha_{01}+\alpha_{10}}{\sqrt{2}}\right|^2$$

$$P[|\psi^-\rangle] = \left|\frac{\alpha_{01}-\alpha_{10}}{\sqrt{2}}\right|^2$$

This Bell state measurement is a very useful and fundamental tool in many quantum communication protocols including teleportation, entanglement swapping etc. The details on how entanglement and Bell State measurement help achieve this will be explained later in this paper.

### C. CHSH Inequality

A very important mathematical representation of entanglement in quantum mechanics is given by the CHSH Inequality [39] [40]. This inequality states that for two classical random variables $A$ and $B$, that can take values $\pm 1$, the following expression holds:

$$S = \left|\mathbb{E}[A.B] + \mathbb{E}[A.\overline{B}] + \mathbb{E}[\overline{A}.B] - \mathbb{E}[\overline{A}.\overline{B}]\right| \leq 2 \quad (22)$$

Here $\mathbb{E}[X]$ denotes the expected value of random variable $X$.

Now, considering a quantum system, where there is an entangled state $\psi$ between two entities A and B; and the possible measurement outcomes are represented by $A, B, \overline{A}$ and $\overline{B}$, then the expected value of an observable is given by:

$$\langle AB \rangle = \langle \psi|A \otimes B|\psi \rangle$$

However, in this situation, there are scenarios where the CHSH inequality given by Eqn. (22) is violated. To be more specific, here the bounds of the inequality become:

$$S \leq 2\sqrt{2} \quad (23)$$

To understand this, let us take an example, where $A = Z$, $B = \frac{1}{\sqrt{2}}(Z-X)$, $\overline{A} = X$ and $\overline{B} = \frac{1}{\sqrt{2}}(Z+X)$. Then the value of $S$ becomes equal to $2\sqrt{2}$.

Thus, CHSH can be used to quantify the entanglement between $A$ and $B$. In other words, higher the value of $S$, higher is the entanglement. The extreme case for which $S = 2\sqrt{2}$, the entanglement is maximum, and we call the state as the maximally entangled state. The application of this concept will be demonstrated in Section VII while introducing the protocols for quantum communications.

### D. Multi-partite Entanglement

Multi-partite entanglement [41] [42] refers to the scenario when the entanglement is shared among more than two parties. In the context of quantum communication, this refers to the scenario when the entanglement is established among three or more network nodes. For establishing multi-partite entanglement, we need an $N$-qubit system, where $N$ is the number of nodes among which the entanglement is shared.

Here, the number of basis states would be equal to $2^N$.

Let us consider a tri-partite entanglement system ($N = 3$). Any general state $|\psi\rangle$ in this three-qubit system can be expressed in terms of the orthogonal basis states:

$$|\psi\rangle = c_1|000\rangle + c_2|001\rangle + c_3|010\rangle + c_4|011\rangle + c_5|100\rangle + c_6|101\rangle + c_7|110\rangle + c_8|111\rangle$$

Two examples of tripartite entangled states are (a) Greenberger-Horne-Zeilinger (GHZ) state [43] and (b) W state [44]. The entangled GHZ state is the equal superposition between the states $|000\rangle$ and $|111\rangle$ and can be expressed as:

$$|GHZ\rangle = \frac{1}{\sqrt{2}}(|000\rangle + |111\rangle) \quad (24)$$

Note that on performing Z-basis measurement on one of the qubits of this state, we obtain the outcome of $+1$ and $-1$ with equal probabilities, that is,

$$P\{+1\} = P\{-1\} = \frac{1}{2} \quad (25)$$

Also, note that, since the qubits are correlated with each other, measurement of one qubit destroys the entanglement. In other words, when the measurement is done on any of the qubits, the entangled state collapses for the entire system. When the measurement outcome is $+1$, the state collapses to $|000\rangle$ and when the outcome is $-1$, the state collapses to $|111\rangle$.

On the other hand, the entangled W state is the equal superposition between three basis states $|001\rangle, |010\rangle$ and $|100\rangle$. The state W can be expressed as:

$$|W\rangle = \frac{1}{\sqrt{3}}(|001\rangle + |010\rangle + |100\rangle) \quad (26)$$

The Z-basis measurement outcomes on the first qubits are as follows:

$$P\{+1\} = \frac{2}{3}$$
$$P\{-1\} = \frac{1}{3}$$

When the measurement outcome is $+1$, the $W$ state now becomes $|0\rangle|\psi^+\rangle$. And when the outcome is $-1$, the state collapses to $|100\rangle$. Thus, unlike the GHZ, the entanglement is not always destroyed for the $W$ state.

In simpler terms, the GHZ state demonstrates the entanglement among all three qubits; whereas the W state demonstrates the case where one qubit is entangled with the other two. In fact, there is a trade-off in the amount of entanglement that can be established in an $N$-qubit system. This is dictated by the principle of monogamy of entanglement [45] [46] [47] that states that when two particles are highly entangled with each other, they cannot be highly entangled with a third particle simultaneously. This is the principle that most of the protocols for quantum communication relies on for security.

The above concept can be easily extended for an $N$-qubit system, where any general state can be expressed as:

$$|\psi\rangle = c_1|00\ldots.0\rangle + c_2|00\ldots.1\rangle \ldots\ldots\ldots + c_{2^N}|1\ldots..11\rangle$$

The GHZ state and W state in this N-qubit system become:

$$|GHZ\rangle = \frac{1}{\sqrt{2}}(|00\ldots\ldots 0\rangle + |11\ldots\ldots 1\rangle) \quad (27)$$

$$|W\rangle = \frac{1}{\sqrt{N}}(|00\ldots\ldots 01\rangle + |00\ldots\ldots 10\rangle + |10\ldots\ldots 00\rangle) \quad (28)$$

## VI. REAL WORLD REALIZATION OF QUANTUM INFORMATION PROCESSING

Till this point in this paper, we have developed the theory and concepts of quantum computing and quantum mechanics. We have seen how quantum states are different than the states obtained from classical bits; how to operate, measure and represent them. The fundamental question that we have not addressed till now is how to create these qubits and quantum states in practice and what these quantum states actually look like in the real world. In other words, in classical communication, the bits '0' and '1' can be encoded by using low and high voltage signals using an electric circuit. We need to answer what is the quantum counterpart of this approach, along with many other related questions.

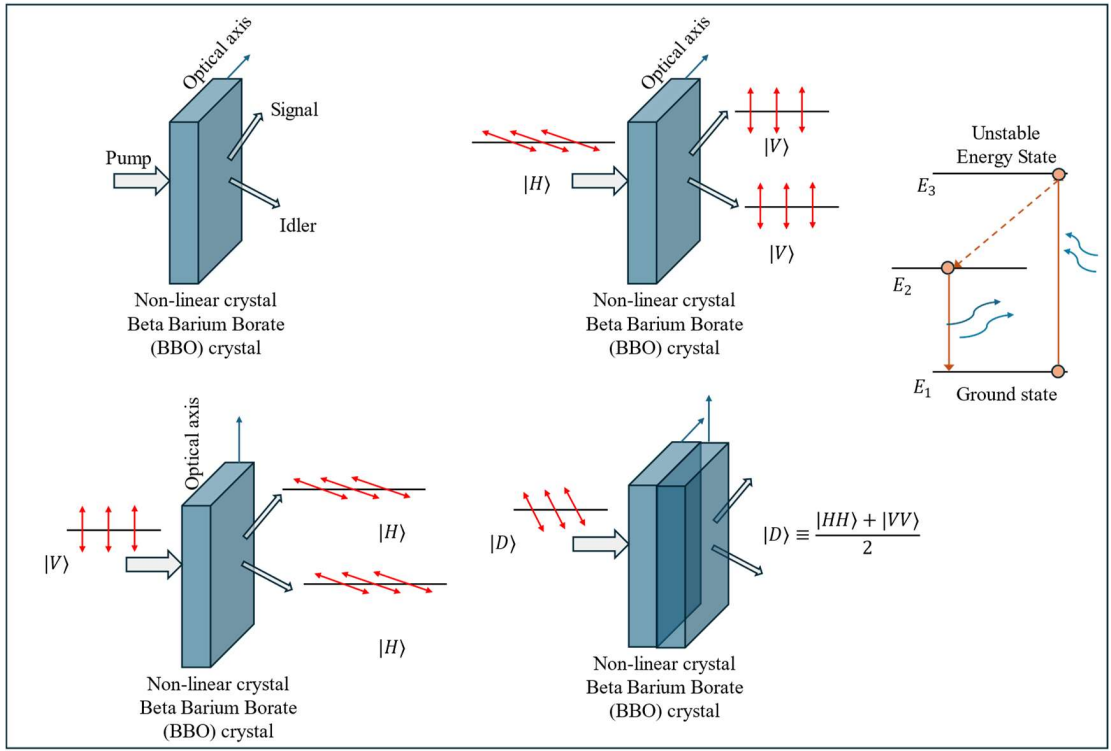

Fig 4: Real world implementation of SPDC

The most common way of realizing quantum communication is by means of photons, since they satisfy the fundamental requirements of quantum states, such as, having the ability of being entangled and in superposed state. Note that single photon states act as the quantum states of light. The challenge here is to develop mechanisms that can generate single photons. However, there are techniques of generating a single photon in the laboratory. One of the popular methods of generating single photons is the Spontaneous Parametric Down Conversion (SPDC) technique [48] [49] [50] [51].

SPDC mechanism can be achieved using a setup as depicted in Fig. 4, where a laser light is made to be incident on a non-linear crystal, such as Beta-Barium Borate (BBO) [52]. The frequency of the incident pump lasing light is adjusted such that it resonates with the frequency of the atomic particles inside the crystal. Now, if we visualize the energy states of the atoms inside the nonlinear crystal, some atoms will get excited from a low (ground) energy state to a high (excited) energy state. Note that the high energy state is an unstable state, so the atoms will come back to the ground state by releasing two photons. These two photons are called 'signal' and 'idler' respectively. By the principle of conservation of energy, the sum of energy of the idler and signal equals the lasing energy used for exciting the atom. Thus, at a macroscopic level, by making a 'frequency-tuned' light beam incident on a non-linear crystal, we can generate two beams called signal and idler.

This approach can be leveraged to encode qubits by using linear polarized incident light. In other words, by making the light oscillate in particular plane, the different quantum states can be generated. To exemplify, light allowed to oscillate in horizontal or vertical plane can be used to represent qubit $|0\rangle$ and $|1\rangle$ respectively. Then the diagonally polarized light would represent an equal superposition of $|0\rangle$ and $|1\rangle$, that is the $|+\rangle$ state. Now depending on the orientation of the nonlinear crystal, that is based on its optical axis, the polarization of the output idler and signal light will vary for a given polarized light. As an example, as shown in Fig. 4, for an incident horizontally polarized light on a specific orientation of the crystal optical axis, the signal and idler would be both vertically polarized. Similarly, if the crystal is rotated and the incident beam is vertically polarized, two horizontally polarized light beams can be obtained. Now if we use two crystals with two different orientations together and allow diagonally polarized light to be incident on it, we get a light that is superposition of horizontally and vertically polarized photon. Thus, following the jargon used a-priori in this paper, polarized light beams, representing $|0\rangle$ and $|1\rangle$, incident on certain orientation of the nonlinear crystal can generate multi-qubit states $|00\rangle$ and $|11\rangle$ respectively. Similarly, a $|+\rangle$ state (represented by a diagonally polarized light) can be used to generate an entangled state $\frac{|00\rangle+|11\rangle}{\sqrt{2}}$.

This process of Spontaneous Parametric Down Conversion (SPDC) can generate one in $10^6$ single photons using the existing technology [50] [51]. However, to be noted that this method, although is able to produce single photons, the generation is still stochastic. For practical implementations, we need a mechanism that can generate single photons deterministically.

Single photons can be generated deterministically in materials with three-level energy structure as shown in Fig. 4. An application of external energy can excite the atom from level $E_1$ to level $E_3$. However, the level $E_3$ is unstable and the atom immediately goes to the stable $E_2$. The level $E_2$ is long-lived enough such that the pulse that is used to excite the atom to level $E_3$ is finished. The atom then transitions from $E_2$ to $E_1$ by releasing a photon. The fact that the intermediate level $E_2$ is so stable gives a distinct amount of time between the end of excitation pulse and the release of the photon due to de-excitation from $E_2$ to $E_1$. This makes the generation event of single photon nearly deterministic. An example of materials that exhibit such properties is Nitrogen-vacancy centers in diamond [53], which is prepared by replacing one carbon atom in diamond with a nitrogen atom and removing another atom to create a vacancy. Such materials are widely used in quantum communication and quantum information processing.

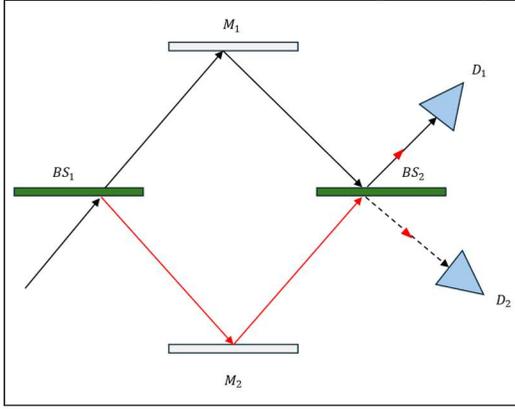

Fig. 5: Mach-Zender Interferometer

Once the qubits are generated, we need some mechanism to implement systems to process these qubits. In other words, we need some systems for realizing different operations on these qubits. One of the most widely used systems to implement quantum information processing is the Mach-Zender Interferometer. The simplistic block diagram for a Mach-Zender Interferometer [54] [55] is shown in Fig. 5 which is composed of two beam-splitters [56] ($BS_1$ and $BS_2$), two mirrors ($M_1$ and $M_2$) and two detectors [57] ($D_1$ and $D_2$). The positions of all these elements in the system are such that the path lengths of the photons across all the detectors are the same. The coding scheme followed here is such that if the photon lies on the bottom half of the interferometer, we call it state $|1\rangle$ and if it lies on the top half, we call it state $|0\rangle$. Now, as an example, let us consider a photon coming from the lower half (that is, it is in state $|1\rangle$) and it hits $BS_1$ from the bottom. Now, that photon can be either reflected back or transmitted through $BS_1$. If it transmits through $BS_1$, it gets reflected by $M_1$ and hits $BS_2$ from the top. Similarly, for the other path, the photon can be reflected by $M_2$ and it reaches $BS_2$ from the bottom. Now, the observation that we make here is based on which detector detects the photon. If detector $D_1$ clicks, then it is $|0\rangle$, and if detector $D_2$ clicks we call it $|1\rangle$. Now experimentally, what we observe is something interesting. If the photon is in state $|1\rangle$, we always observe a click on $D_1$, that is, observation is $|0\rangle$. Otherwise, when the photon is in state $|0\rangle$, we always observe a click on $D_1$, that is, observation is $|1\rangle$. This is an example implementation of the NOT gate.

Mathematically, we can think of the two beam splitters as:
$$BS_1 = \frac{1}{\sqrt{2}}\begin{bmatrix} 1 & 1 \\ 1 & -1 \end{bmatrix} \text{ and } BS_2 = \frac{1}{\sqrt{2}}\begin{bmatrix} -1 & 1 \\ 1 & 1 \end{bmatrix}$$

For an input photon in state $|\phi\rangle = |1\rangle$, the operations can be visualized as:

$$BS_2 . BS_1 |\phi\rangle$$
$$= BS_2 \frac{1}{\sqrt{2}}\begin{bmatrix} 1 & 1 \\ 1 & -1 \end{bmatrix}\begin{bmatrix} 0 \\ 1 \end{bmatrix}$$
$$= BS_2 \frac{1}{\sqrt{2}}\begin{bmatrix} 1 \\ -1 \end{bmatrix}$$
$$= \frac{1}{\sqrt{2}}\begin{bmatrix} -1 & 1 \\ 1 & 1 \end{bmatrix} . \frac{1}{\sqrt{2}}\begin{bmatrix} 1 \\ -1 \end{bmatrix}$$
$$= \frac{1}{2}\begin{bmatrix} -2 \\ 0 \end{bmatrix} = \begin{bmatrix} -1 \\ 0 \end{bmatrix} = -|0\rangle = |0\rangle$$

Similarly, an input photon in state $|\phi\rangle = |0\rangle$, the operations can be visualized as:

$$BS_2 . BS_1 |\phi\rangle$$
$$= BS_2 \frac{1}{\sqrt{2}}\begin{bmatrix} 1 & 1 \\ 1 & -1 \end{bmatrix}\begin{bmatrix} 1 \\ 0 \end{bmatrix}$$
$$= BS_2 \frac{1}{\sqrt{2}}\begin{bmatrix} 1 \\ 1 \end{bmatrix}$$
$$= \frac{1}{\sqrt{2}}\begin{bmatrix} -1 & 1 \\ 1 & 1 \end{bmatrix} . \frac{1}{\sqrt{2}}\begin{bmatrix} 1 \\ 1 \end{bmatrix}$$
$$= \frac{1}{2}\begin{bmatrix} 0 \\ 2 \end{bmatrix} = \begin{bmatrix} 0 \\ 1 \end{bmatrix} = |1\rangle$$

Next, in the interferometer, if we block the photon to reach the mirror $M_2$ by putting a light absorbing material, then the interference at beam splitter $BS_2$ is prevented (Fig. 5). The observations here are as follows. For some of the experimental runs, we see the detector $D_1$ click and for some other runs, we see $D_2$ click. Statistically, if we run experiments long enough, we see that the probability of each detector clicking is the same. In the jargon of quantum mechanics, what we achieve in this scenario is an equal superposition of states $|1\rangle$ and $|0\rangle$. In other words, this phenomenon is equivalent to a Hadamard transformation:

$$|0\rangle \xrightarrow{H} \frac{|0\rangle+|1\rangle}{\sqrt{2}} \text{ and } |1\rangle \xrightarrow{H} \frac{|0\rangle-|1\rangle}{\sqrt{2}} \quad (29)$$

VII. ESTABLISHMENT OF QUANTUM COMMUNICATION

This section talks about the procedure of establishing quantum communication over a link. We will first discuss how quantum communication can be established between two nodes. For this, we will introduce the phenomenon of teleportation, the fundamental characteristics of quantum communication. It would be shown how teleportation leverages the concept of entanglement to establish secure communication between two nodes. This leads to the explanation of No Cloning Theorem and its significance in teleportation. Next, we will introduce some secure quantum communication protocols, which comprise of the core of this communication technology today.

A. Teleportation

To understand the teleportation protocol [58] [59] [60], let us consider two nodes A and B, where A wants to share some information to B. Node A has an arbitrary qubit $|\psi\rangle_{A_1} = \alpha|0\rangle +$

$\beta|1\rangle$). It also shares an entangled state with node B. Let us assume that the shared entangled state is $|\phi^+\rangle_{A_2B} = \frac{1}{\sqrt{2}}(|00\rangle + |11\rangle)$. Thus, the total initial state of the system can be represented as:

$$|\psi\rangle_{A_1}|\phi^+\rangle_{A_2B} = (\alpha|0\rangle + \beta|1\rangle)(\frac{1}{\sqrt{2}}(|00\rangle + |11\rangle))$$

$$= \frac{1}{\sqrt{2}}(\alpha|000\rangle + \alpha|011\rangle + \beta|100\rangle + \beta|111\rangle)$$

$$= \frac{1}{\sqrt{2}}(\alpha(\frac{1}{\sqrt{2}}(|\phi^+\rangle + |\phi^-\rangle))|0\rangle + \alpha(\frac{1}{\sqrt{2}}(|\psi^+\rangle + |\psi^-\rangle))|1\rangle$$

$$+ \beta(\frac{1}{\sqrt{2}}(|\psi^+\rangle - |\psi^-\rangle))|0\rangle + \beta(\frac{1}{\sqrt{2}}(|\phi^+\rangle - |\phi^-\rangle))|1\rangle)$$

$$= \frac{1}{2}((\alpha|0\rangle + \beta|1\rangle)|\phi^+\rangle + (\alpha|0\rangle - \beta|1\rangle)|\phi^-\rangle + (\alpha|1\rangle + \beta|0\rangle)|\psi^+\rangle + (\alpha|1\rangle - \beta|0\rangle)|\psi^-\rangle) \quad (30)$$

Note that the probability of obtaining all the possible bell states measurements $|\phi^+\rangle, |\phi^-\rangle, |\psi^+\rangle, |\psi^-\rangle$ are equal. Next, node A performs measurement on this initial state, where all the measurements are equally likely. Node A shares the outcome of the measurement to node B, using classical communication of two bits (since it is a two-qubit state). Now, let us analyze the measurement outcome of A.

*Scenario 1*: Node A's measurement outcome is $|\phi^+\rangle$. This means that B has the state $\alpha|0\rangle + \beta|1\rangle$, which is the desired information that A wants to send to B.

*Scenario 2*: When A measures $|\phi^-\rangle$, this indicates that B has the state $\alpha|0\rangle - \beta|1\rangle$. In this case, when B knows that A measured $|\phi^-\rangle$, it performs a unitary operation $Z$ on its quantum state to know what A wanted to communicate. Mathematically:

$$Z(\alpha|0\rangle - \beta|1\rangle) = \begin{bmatrix}1 & 0\\0 & -1\end{bmatrix}(\alpha|0\rangle - \beta|1\rangle)$$

$$= \begin{bmatrix}1 & 0\\0 & -1\end{bmatrix}(\alpha\begin{bmatrix}1\\0\end{bmatrix} - \beta\begin{bmatrix}0\\1\end{bmatrix})$$

$$= (\alpha\begin{bmatrix}1\\0\end{bmatrix} - \beta\begin{bmatrix}0\\-1\end{bmatrix})$$

$$= \alpha|0\rangle + \beta|1\rangle$$

Thus, we can observe that B can retrieve the information that A wants to share, resulting in successful teleportation.

*Scenario 3*: When the measurement outcome of A is $|\psi^+\rangle$, B performs the unitary operation of $X$ in its state. Note that, in this measurement, node B has the state $\alpha|1\rangle + \beta|0\rangle$. This results in the following operation in Z's state:

$$X(\alpha|1\rangle + \beta|0\rangle) = \begin{bmatrix}0 & 1\\1 & 0\end{bmatrix}(\alpha|1\rangle + \beta|0\rangle)$$

$$= \begin{bmatrix}0 & 1\\1 & 0\end{bmatrix}(\alpha\begin{bmatrix}0\\1\end{bmatrix} + \beta\begin{bmatrix}1\\0\end{bmatrix})$$

$$= (\alpha\begin{bmatrix}1\\0\end{bmatrix} + \beta\begin{bmatrix}0\\1\end{bmatrix})$$

$$= \alpha|0\rangle + \beta|1\rangle$$

In this way, node B knows the information that node A wants to share.

*Scenario 4*: The final possibility is that node A measures the state as $|\psi^-\rangle$, when the state of node B is $\alpha|1\rangle - \beta|0\rangle$. In this case, node B does the unitary operation of $ZX$ to find out the information shared by node A as follows.

$$ZX(\alpha|1\rangle - \beta|0\rangle) = Z\begin{bmatrix}0 & 1\\1 & 0\end{bmatrix}(\alpha|1\rangle - \beta|0\rangle)$$

$$= Z\begin{bmatrix}0 & 1\\1 & 0\end{bmatrix}(\alpha\begin{bmatrix}0\\1\end{bmatrix} - \beta\begin{bmatrix}1\\0\end{bmatrix})$$

$$= Z(\alpha\begin{bmatrix}1\\0\end{bmatrix} - \beta\begin{bmatrix}0\\1\end{bmatrix})$$

$$= \begin{bmatrix}1 & 0\\0 & -1\end{bmatrix}(\alpha\begin{bmatrix}1\\0\end{bmatrix} - \beta\begin{bmatrix}0\\1\end{bmatrix})$$

$$= \alpha|0\rangle + \beta|1\rangle$$

Thus, irrespective of the equally likely measurement outcomes of A, node B can figure out the information that node A wants to communicate. This protocol used in quantum communication is known as teleportation.

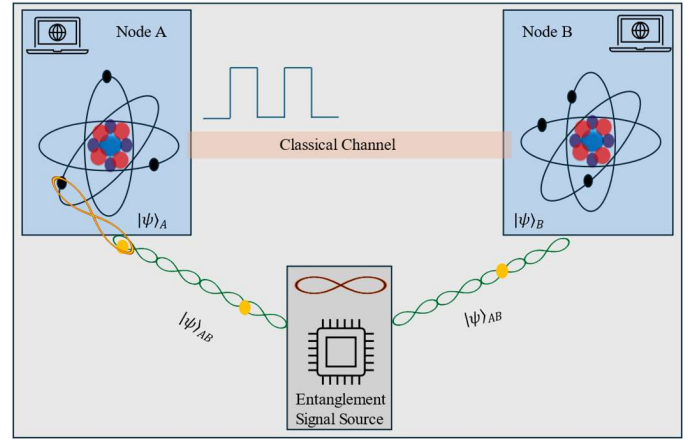

Fig. 6: Visual Representation of teleportation between two nodes

There are several key points to consider about the teleportation protocol:

1. The teleportation protocol relies on the establishment of successful entanglement between the two nodes participating in communication. Without successful entanglement, node B would never be able to know what node A wanted to communicate.

2. Any arbitrary node in the network who does not share entanglement with node A will never be able to retrieve information that A wants to send to B. What A sends to B is the outcome of measurement on the initial state. This information has no meaning unless there is a successful entanglement between the two parties in communication.

3. Initially, before A performs the measurement and shares the outcome of it, state of A is the one that A wants to communicate to B (that is state $|\psi\rangle$) and state of node B is in maximally mixed state. After the teleportation protocol, node A projects its qubits to one of the four possible Bell states, where all of them are maximally mixed and node B's qubit becomes state $|\psi\rangle$.

4. The sharing of the measurement outcome for successful execution of teleportation is done using classical bits. In other words, this teleportation protocol can be thought of as a higher layer of operation for achievement of secure communication.

5. The teleportation that we have seen in the example above relies on entanglement between two nodes. This is called Bipartite entanglement. When communication has to be performed between more than two nodes, the entanglement has also to be done among all these nodes, which is the multipartite entanglement, discussed earlier.

6. The teleportation protocol does not violate the spatial theory of relativity. In other words, teleportation is not instantaneous, and it is not faster than light communication. To exemplify, initially, at time $t_0$, node A and B have a shared entangled state, when node B's qubit is in maximally mixed state. At time $t_1 > t_0$, node A does measurement in the Bell basis and shares the measurement outcome with node B. Even at this time instant, node B's qubit is in maximally mixed state. When node B receives the measurement outcome from node A at time $t_2 > t_1$, node B knows the state of node A, by performing a unitary operation on its state. This is the time when teleportation has successfully been taken place between A and B and as seen above there is a time $(t_2 - t_0)$ required for its execution.

### B. No Cloning Theorem

One of the fundamental and most widely used concepts in quantum communication is the 'No Cloning Theorem.' [61] [62] [63] This theorem from quantum mechanics states that it is impossible to recreate or clone any arbitrary, unknown quantum state $|\psi\rangle$. Mathematically, it can be formulated as, for an arbitrary state $|\psi\rangle$, if there exists an operator $U$ that can clone state $|\psi\rangle$ to state $|\phi\rangle$, such that $U|\psi\rangle|0\rangle = |\psi\rangle|\psi\rangle$, where $|0\rangle$ is the initial blank state, then this would violate the unitary and linearity of quantum mechanics.

This can be proved by considering a state $|\Psi\rangle$ that is a superposition of states $|\psi\rangle$ and $|\phi\rangle$. That is,

$$|\Psi\rangle = \alpha|\psi\rangle + \beta|\phi\rangle$$

Now let us assume there exists an operator $U$ that can clone state $|\Psi\rangle$.

$$U|\Psi\rangle|0\rangle = |\Psi\rangle|\Psi\rangle \quad (31)$$

Now replacing $|\Psi\rangle = \alpha|\psi\rangle + \beta|\phi\rangle$ in left hand side of Eq. (31), we get

$$U|\Psi\rangle|0\rangle = U(\alpha|\psi\rangle + \beta|\phi\rangle)|0\rangle$$
$$= \alpha U|\psi\rangle|0\rangle + \beta U|\phi\rangle|0\rangle$$
$$= \alpha|\psi\rangle|\psi\rangle + \beta|\phi\rangle|\phi\rangle \quad (32a)$$

Using $|\Psi\rangle = \alpha|\psi\rangle + \beta|\phi\rangle$ in right hand side of Eq. (32):
$$|\Psi\rangle|\Psi\rangle = (\alpha|\psi\rangle + \beta|\phi\rangle)(\alpha|\psi\rangle + \beta|\phi\rangle))$$
$$= \alpha^2|\psi\rangle|\psi\rangle + \beta^2|\phi\rangle|\phi\rangle + \alpha\beta|\psi\rangle|\phi\rangle + \alpha\beta|\phi\rangle|\psi\rangle \quad (32b)$$

Comparing Eqns. (32 a) and (32 b), we see that the terms $|\psi\rangle|\phi\rangle$ and $|\phi\rangle|\psi\rangle$ from (32 b) do not appear in (32 a). This is contradictory to the linearity of quantum mechanics. Thus, it is not possible to create an identical copy or clone any arbitrary quantum state.

This theorem holds significance from the quantum information and computing perspective. In classical computing, any information can be recreated or cloned, which cannot be done in quantum computing, as suggested by the '*No Cloning Theorem.*' Quantum computing algorithms often rely on manipulating and processing quantum states and qubits. The '*No Cloning Theorem*' implies that the quantum computers cannot perform 'copy-like' processing on the quantum information. This unique and strong characteristic of quantum computing has been the root of many fundamental protocols of quantum communication and cryptography. This '*No Cloning Theorem*' makes sure secure execution of Quantum Key Distribution (QKD) protocol [64] [65] [66], that does not allow eavesdroppers to clone and recreate quantum information without getting noticed.

### C. Protocols for Secure Quantum Communications

The standard cryptographic techniques involved in classical communication are public key cryptography [67] and private key cryptography [68]. To get a background let us understand this in the setup of a node-to-node communication between Alice and Bob, where Bob wants to send a message to Alice via a public channel. In public key cryptography, Alice shares a public key with Bob over the public channel. Bob receives the public key, using which, it encrypts its message and then sends the encrypted message to Alice. Upon receiving the encrypted message, Alice decrypts it using a private key that it did not share with anyone. In other words, in public key cryptography, the public key is used to encrypt the message and the private key is used to decrypt it. Public key cryptography is slow, expensive and computationally secure. That is, the encryption can be broken, given sufficiently large computational resources. The other classical cryptography technique, private key cryptography uses a private encryption key to be shared between the transmitter and receiver using a separate private channel. This is usually done by a central arbitrator that shares the private key between them, that creates physical realizability problem. This private key cryptography is secure until and unless the key is used only once. This technique also requires that the length of the key has to be at least the size of the message, which leads to scalability issues.

### i) Quantum Key Distribution (QKD) Protocols

There are cryptographic protocols used in quantum communication for solving several of the issues of these classical techniques. One of the most fundamental protocols used for security in quantum communication is the BB84 protocol [69] [70] [71] proposed by Bennett and Brassard in 1984. To understand this protocol let us assume the same scenario for node-to-node communication between Alice and Bob. At the start of the protocol, Alice generates two random bit-strings $\{a_1, a_2, \ldots, a_n\}$ and $\{b_1, b_2, \ldots, b_n\}$. Now, Alice prepares a two-qubit state sequence $\{|\psi\rangle_1, |\psi\rangle_2, \ldots, |\psi\rangle_n\}$ using these two bit-strings as follows:

$$|\psi\rangle_i = |\psi_{a_i b_i}\rangle \quad (33)$$

Here, $|\psi_{00}\rangle = |0\rangle, |\psi_{01}\rangle = |+\rangle, |\psi_{10}\rangle = |1\rangle, |\psi_{10}\rangle = |-\rangle$

Note that the bit-string $\{b_i\}$ decides the basis of preparation of the qubit. If $b_i = 1$, the preparation basis is $Z$ and if $b_i = 0$, the basis is $X$. Another point to pay attention here is that the qubit pairs $(|\psi_{00}\rangle, |\psi_{01}\rangle)$ and $(|\psi_{10}\rangle, |\psi_{11}\rangle)$ not orthogonal and hence they are not distinguishable. That is,

$$\langle\psi_{00}|\psi_{01}\rangle = \langle\psi_{10}|\psi_{11}\rangle = \frac{1}{\sqrt{2}} \quad (34)$$

In other words, if the same-basis measurement is done on these states, they won't result in deterministic results, unlike the case for measurement in orthogonal states, like $(|0\rangle, |1\rangle)$, where the measurement in $Z$ basis would result in +1 and -1 outcome respectively. This non-distinguishable feature is made sure by using the randomness in the bit-string of $\{b_i\}$.

Next Alice sends the strings of qubits $\{|\psi\rangle_i\}$ over the quantum channel. Bob does not have the knowledge of the basis of preparation of the qubits till this point, since Alice has not shared the two strings with Bob yet. Upon receiving the quantum states, Bob generates a random bit-string $\{b_1', b_2', \ldots, b_n'\}$. Now based on the value of $b_i$, Bob decides the basis of measurement for the received quantum state $|\psi\rangle_i$. That is, if $b_i' = 1$, the measurement basis is $Z$ and if $b_i' = 0$, the basis is $X$. If the measurement outcome for $|\psi\rangle_i$ is +1, then Bob stores $a_i' = 1$ and if measurement outcome is -1, then $a_i' = 0$. After the measurement operation, Alice and Bob share the strings $\{b_i\}$ and $\{b_i'\}$ with each other and they keep $a_i$ and $a_i'$ for which $b_i = b_i'$. This results in new shorter keys $\hat{A} = \{\hat{a}_1, \hat{a}_2, \ldots, \hat{a}_m\}$ ($m \leq n$) such that $\hat{a}_1 = \hat{a}_1'$.

Now, let us understand the scenario when there is an eavesdropper in the network. The simplistic approach for an eavesdropper is to copy the quantum states sent from Alice to Bob, and then send it again to Bob. But fortunately, this is not possible because of the '*No-Cloning Theorem*' stated above, which says that quantum states cannot be cloned or replicated. Thus, the only way for the eve to know the information sent from Alice to Bob is to measure the quantum state. In this case, the correct measurement is possible if the eavesdropper knows the preparation basis, which is not the case here. Therefore, the eve randomly guesses the measurement basis and performs the measurement. However, if the measurement basis is not the same as the preparation basis, it will create disturbance in the quantum states, and this will disrupt the sent quantum information. This would allow Alice and Bob to know the presence of an eavesdropper on the channel.

In order to detect an eavesdropper, Alice and Bob would assign a part of their shared key $\hat{A}$ for this purpose. Any discrepancy in those bits suggest that there is a high possibility of the presence of an eavesdropping agent.

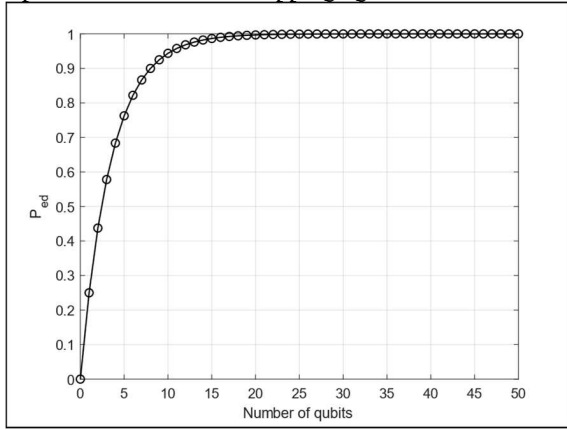

Fig. 7: Probability of eavesdropper detection

Another approach is to detect eavesdroppers on the go, instead of allocating separate bits. Here, the same idea is used, that is, if the measurement basis for the eve is different than the preparation basis for Alice and the measurement basis of Bob is the same as that of Alice, then the eavesdropper can be detected due to disturbance in the qubits. Note that with an increase in the number of qubits in the system, the probability of eavesdropper detection increases. For a single qubit, two-basis system, the probability that the eve will use a different basis than Alice is $\frac{1}{2}$ and that Bob will use the same basis as Alice is $\frac{1}{2}$. Thus, for an n-qubit system, the probability of eavesdropper detection can be expressed as:

$$P_{ed}(n) = 1 - \left[1 - \left(\frac{1}{2}\right)\left(\frac{1}{2}\right)\right]^n = 1 - \left(\frac{3}{4}\right)^n \quad (35)$$

The dependence of the eavesdropper detection probability on number of qubits is demonstrated in Fig. 7. From the figure, it can be observed that for a 20-qubit system the probability of eavesdropper detection is 99.68% and for a 30-qubit system, this probability becomes 99.98%.

The first real world quantum network implementing QKD protocol was built by a team from BBN Technologies, Harvard, and Boston University in 2003 [72]. This network, known as DARPA quantum network, expands in the Massachusetts area over a length of 29 km in total. It demonstrated the viability of quantum communications over different physical links, including free-space and SMF-28 telecommunications fiber [73]. Fig. 8 gives an overview of the communication framework developed by them, where nodes 'Ali' and 'Baba' run the BBN QKD protocols using a free-space communication. These two nodes are linked into the rest of the network by BBN's QKD key relay protocols between nodes 'Ali' and 'Alice'. The paper reported that the mean photon number of the node 'Anna' was 0.5. The Anna-Bob system delivered about 1,000 privacy-amplified secret bits/second at an average 3% Quantum Bit Error Rate (QBER).

The second implementation of the QKD-based network was in Vienna in the year 2008 [74]. This network was called SEcure COmmunication based on Quantum Cryptography (SECOQC) and it was developed for an eight point-to-point links, six-nodes network. The reported longest link length was 83 km and the order of magnitude of secret transmission capacity of the prototype was in the range of 1 GB per month.

There have been multiple attempts to develop and improve performance of the quantum networks using secure QKD-protocol. A dynamically reconfigurable network was developed in a testbed of the Advanced Technology Demonstration Network (ATDNet) by Telcordia Technologies [75]. A hierarchical network consisting of a 5-node wavelength division multiplexing (WDM) quantum backbone and subnets connected by trusted nodes and an all pass optical switching network were deployed in China [76], [77]. Another significant development was reported in paper [78] where QKD-based protocol was implemented to demonstrate video conferencing in Japan in 2010. This network had four access points, Koganei, Otemachi, Hakusan, and Hongothat were connected by a bundle of commercial fibers.

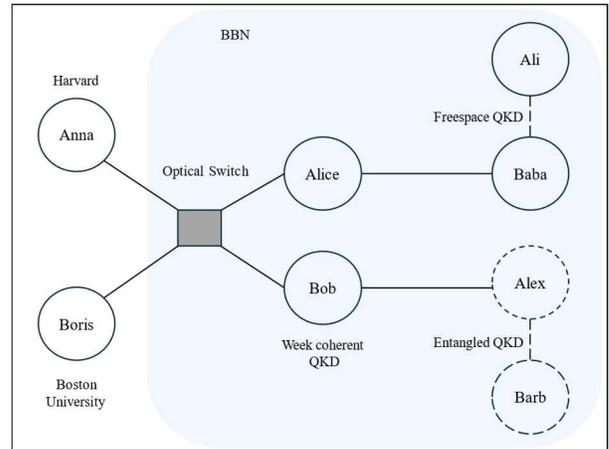

Fig. 8: Schematic of DARPA quantum network

*ii) Entanglement-based Quantum Key Distribution Protocols*

Let us first understand the limitations of the QKD-based BB84 protocol that we discussed in the last section. The BB84 protocol is secure as long as the eavesdropper does not know the preparation basis of the quantum states of Alice. However, if there exists a scenario where the eavesdropper has a mechanism to correctly predict the preparation basis of Alice, the protocol is no longer secure. This limitation can be handled using Entangled-based QKD protocols as the E91 protocol [79] [80] [81] [82], which we are going to explain next.

The main idea of E91 protocol is the property of "Monogamy of Entanglement," which states that as the entanglement between Alice and Bob becomes stronger, its correlation with a third party becomes less probable. In the extreme case, for a maximally entangled state, it cannot be correlated with the third part, such as, an eavesdropper, and as a result, the communication becomes secure. The E91 protocol leverages this property by making the communicating nodes to make sure that the entanglement between them is strong enough.

In this protocol, Alice and Bob use three bases for preparation and measurement of the quantum states. The bases for Alice's measurement are $A_1 = Z, A_2 = X, A_3 = \frac{1}{\sqrt{2}}(Z + X)$. Similarly, the bases for Bob's measurement are $B_1 = Z, B_2 = \frac{1}{\sqrt{2}}(Z + X), B_3 = \frac{1}{\sqrt{2}}(Z + X)$. These communicating entities randomly pick a basis for preparation and measurement respectively, similar to BB84 protocol. Alice sends the strings of qubits over the quantum channel. Upon receiving the quantum states, Bob performs the measurement by randomly selecting a basis out of those three bases. Alice and Bob shares the bases used for preparation and they keep the measurement outcomes for which they used the same bases. Till this point, this protocol is very similar to the BB84 protocol. However, the difference comes from the fact that unlike BB84, the communicating entities Alice and Bob do not discard the outcomes for which the bases do not match. They use these to compute the following expression:

$$S = |\langle A_1 B_2 \rangle + \langle A_1 B_3 \rangle + \langle A_2 B_2 \rangle - \langle A_2 B_3 \rangle| \quad (36)$$

Now, from the discussion in Section V (C), we know that this is the expression for CHSH inequality [83]; and the value of this expression should be greater than 2 to ensure entanglement. Therefore, Alice and Bob checks the CHSH expression and if this is less than or equal to 2, they abort the communication, since it indicates that there is no entanglement. And by the property of "Monogamy of Entanglement," there is a possibility of the presence of an eavesdropper. Thus, the key produced using E91 protocol is unconditionally secure, in contrary to BB84 protocol that assumes that the eavesdropper does not know the preparation basis for Alice's qubits.

A real-world experiment of Entanglement-based QKD protocol was conducted between two islands in Canary Islands in the year 2007 [84]. The communication link was free space over a distance of 144 km. The entanglement was done using polarized photons generated using SPDC mechanism as we discussed earlier in this paper. The reported CHSH violation in this paper was $2.508 \pm 0.037$. Another prototype for Entanglement-based QKD protocol was developed in 2016 as reported in the paper [85], that used optical fiber as the communication medium. They demonstrated a bit rate of the order of $10^{-4}$ over a distance of 400 km, using an ultra-low loss fiber. The paper [86] showed the Entanglement-based QKD for satellite communications, which reported a CHSH violation of $2.56 \pm 0.007$ and a bit rate of 0.12. Recently in 2022, demonstration of entanglement-based QKD in 270 m long free-space link in day light was done in [87]. This used a quantum dot source for narrower spectral bandwidth—beneficial for filtering out sunlight—and negligible multiphoton emission at peak brightness to make their system run for continuous three days. The secure data rate obtained in this work was reported to be 12 bps. Wavelength and Space Division Multiplexing was used to develop a network for entanglement based QKD in a fully connected setup of 40 users and 780 QKD links in [88]. It was reported that the average secure key rate between users in the same subnet was ~ 51 bps and that between users in different subnets was ~ 22 bps.

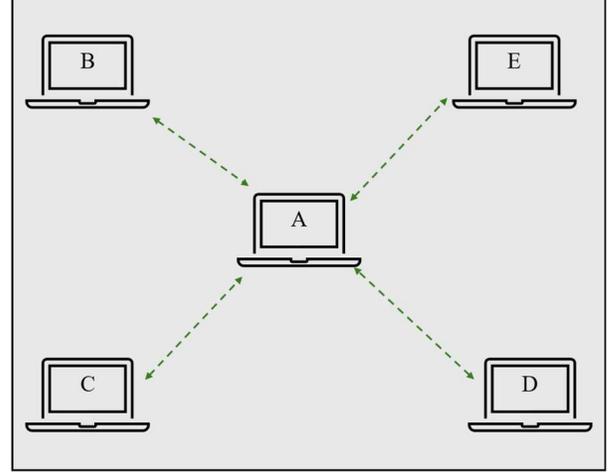

Fig. 9: Distributed quantum computing

### D. Distributed and Blind Quantum Computing

One interesting application of entanglement and quantum communication is in distributed quantum computation, when a device has limited quantum resource [89] [90] [91] [92] [93]. Limitation in quantum resources can be in the form of generation of the number of qubits. In that case, multiple such devices can perform quantum computation by setting up entanglement among them. As shown in Fig. 9, resource-constrained nodes A, B, C, D and E share some small number of qubits, entangle them and then perform quantum operations locally. This ensures the accomplishment of a complex quantum operation using a network of quantum processors with limited quantum resources.

In the scenario of extreme constraints of quantum resources, the nodes can share instructions (in a classical way) to another node with high computation capability for performing the desired quantum computing task. The node *B* with high computation complexity does the computation and shares the result of the operation with the low-complex node *A*. However, this approach is not reliable from the privacy perspective, as the high complexity node knows all the information about the task that the low-complexity node *A* wants to perform. This can be resolved using the concept of blind quantum computation, where the low complex node *A* generates the qubit, rotates it in some basis and then shares the rotated qubit to the high-complexity node *B*. Moreover, node *B* is also informed the operation it needs to perform on a qubit-by-qubit basis. Node *B* performs the assigned task on the rotated qubit and then shares the outcome to the resource-constrained node *A*.

### E. Key Challenges for Quantum Communications

As we have pointed out in the previous discussion of this paper, quantum communication is heavily dependent on entanglement between individual photons. In other words, loss of a photon would have direct implication of failure in quantum communication, unlike classical communication, where loss of a single photon does not bear much significance. To make things more complicated, it is not possible to create backup copies of the photons, as stated by the '*No Cloning Theorem,*' unless we limit ourselves to orthogonal states.

There are several other sources of quantum errors. One such class of errors result from the unitary operations on the quantum states. As an example, phase errors may arise due to operation of unitary $Z$ operation on the quantum states in consideration. There are non-unitary errors as well. One common error in quantum system arises from Decoherence [94], which is the outcome of interaction of quantum states with external environment. Decoherence mainly has two outcomes: loss of probability amplitude (relaxation) and loss of phase information (dephasing). These errors often result in loss of quantum superposition and entanglement over time and destruction of delicate quantum states necessary for quantum communication. A very common outcome of Decoherence is Energy Relaxation or T1 Relaxation [95], where a qubits state to decays from $|1\rangle$ to $|0\rangle$ within a certain amount of time, resulting from an energy flow between spins of quantum particles and their external environment.

In the presence of these errors, it becomes really important to device mechanisms to deal with them. Especially for long distance communication, it is necessary to overcome these quantum challenges, which we are going to present in the next section. In addition, we will study how to establish end-to-end quantum entanglement to facilitate long distance multi-hop quantum communication.

## VIII. LONG DISTANCE QUANTUM COMMUNICATION

In this section, we will leverage the concepts that we have developed till now to extend them for establishing reliable end-to-end long distance quantum communication [96] [97] [98] [99] [100] [101] [102] [103]. We will first demonstrate how entanglement between two nodes that are directly connected and then extend that concept to show how this is achieved for nodes that are multiple hops apart.

### A. Link level Entanglement

There are primarily two approaches for establishing entanglement between two nodes: Midpoint Interference (MIM) and Memory-to-memory (MM) [104] [105] [106] [107] [108].

In Midpoint Interference (MIM) approach, each node is equipped with a certain amount of quantum memory for storing quantum state information. It is demonstrated in Fig. 10, for two neighboring nodes A and B. In order to achieve entanglement between A and B, there is Bell State Analyzer (BSA) [109] [110] placed somewhere in between A and B. Nodes A and B emit photons such that these photons are entangled with the quantum memory of the respective nodes. Multiple such photons are generated in order to account for the loss of photons over the channel. These photons are then transmitted over the channel (optical fiber) and when these photons click one of the detectors of BSA, they are projected to one of the Bell states depending on the measurement outcome. By doing so, the quantum states are also projected to one of the Bell states. To understand this, let us represent the entanglement between the photon and quantum memory in A is given by $|\phi^+\rangle_{AA'} = \frac{1}{\sqrt{2}}(|00\rangle + |11\rangle)$. Similarly, the entangled state for node B is given by $|\phi^-\rangle_{B'B} = \frac{1}{\sqrt{2}}(|00\rangle - |11\rangle)$. So, the total state of the system is given by:

$$|\psi\rangle = |\phi^+\rangle_{AA'}|\phi^-\rangle_{B'B}$$
$$= \frac{1}{2}(|0000\rangle - |0011\rangle + |1100\rangle - |1111\rangle)$$
$$= \frac{1}{2}(|0\rangle|00\rangle|0\rangle - |0\rangle|01\rangle|1\rangle + |1\rangle|10\rangle|0\rangle - |1\rangle|11\rangle|1\rangle)$$
$$= \frac{1}{2\sqrt{2}}(|0\rangle(|\phi^+\rangle + |\phi^-\rangle)|0\rangle - |0\rangle(|\psi^+\rangle + |\psi^-\rangle)|1\rangle$$
$$+ |1\rangle(|\psi^+\rangle - |\psi^-\rangle)|0\rangle - |1\rangle(|\phi^+\rangle - |\phi^-\rangle)|1\rangle)$$
$$= \frac{1}{2\sqrt{2}}((|\phi^+\rangle + |\phi^-\rangle)|0\rangle|0\rangle - (|\psi^+\rangle + |\psi^-\rangle)|0\rangle|1\rangle + (|\psi^+\rangle$$
$$- |\psi^-\rangle)|1\rangle|0\rangle - (|\phi^+\rangle - |\phi^-\rangle)|1\rangle|1\rangle)$$
$$= \frac{1}{2\sqrt{2}}(|\phi^+\rangle(|0\rangle|0\rangle - |1\rangle|1\rangle) + |\phi^-\rangle(|0\rangle|0\rangle + |1\rangle|1\rangle) -$$
$$|\psi^+\rangle(|0\rangle|1\rangle - |1\rangle|0\rangle) - |\psi^+\rangle(|0\rangle|1\rangle + |1\rangle|0\rangle)) \quad (37)$$

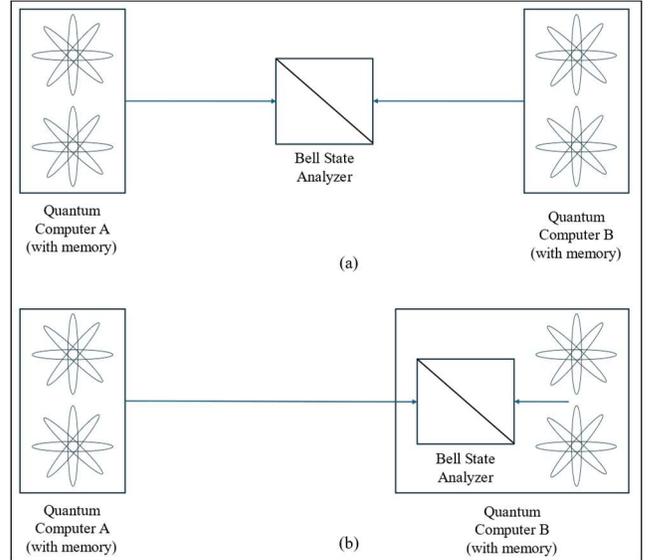

Fig. 10: Establishment of link-level entanglement

Next, upon measurement of the states of the photons $A'B'$, depending on the Bell state ($|\phi^\pm\rangle, |\psi^\pm\rangle$) obtained after measurement, quantum memories that are held at nodes A and B can be projected into one of the maximally entangled Bell states ($|0\rangle|0\rangle \mp |1\rangle|1\rangle, |0\rangle|1\rangle \mp |1\rangle|0\rangle$), as can be seen from the above equation (37).

The setup of Memory-to-memory (MM) approach for entanglement is shown in Fig. 10. The concept of establishing entanglement is the same as in MIM approach. The only difference is that the BSA is located in the node itself instead of keeping it in the middle of the two nodes. The photon is directly sent to the node and the measurement is done inside the node. This approach can be applied in situations where the attenuation of the connecting fibers is low or the distance between the two nodes is small.

### B. Entanglement Swapping

The above concept of establishing link level entanglement can be extended to set up end-to-end entanglement between two nodes that are multi-hops apart [111] [112] [113] [114] [115]. To understand this, let us assume that nodes A and B are connected to node C. Now, we know that link-level entanglement can be set up between A-C and B-C. Next, let us consider, nodes A and C share entanglement that is given by $|\phi^+\rangle_{AA'} = \frac{1}{\sqrt{2}}(|00\rangle + |11\rangle)$, where node A shares qubit $A$ and node C shares qubit $A'$ of the entangled state. Similarly, nodes B and C share entanglement that is given by $|\phi^+\rangle_{BB'} = \frac{1}{\sqrt{2}}(|00\rangle + |11\rangle)$, where node B shares qubit $B$ and node C shares qubit $B'$ of the entangled state. Now, from the discussion of link-level entanglement, the total state of the system is:

$$|\psi\rangle_{AA'B'B} = \frac{1}{2\sqrt{2}}(|\phi^+\rangle(|0\rangle|0\rangle - |1\rangle|1\rangle) + |\phi^-\rangle(|0\rangle|0\rangle + |1\rangle|1\rangle) - |\psi^+\rangle(|0\rangle|1\rangle - |1\rangle|0\rangle) - |\psi^+\rangle(|0\rangle|1\rangle + |1\rangle|0\rangle))$$

Thus, based on the measurement of qubits $A'$ and $B'$ in the intermediate node C, the entanglement between nodes A and B can be set up by projecting into one of the maximally entangled Bell states $(|0\rangle|0\rangle \mp |1\rangle|1\rangle, |0\rangle|1\rangle \mp |1\rangle|0\rangle)$. After measurement is performed at C, nodes A and B are classically notified by C that measurement has been carried out successfully. Node C also lets one of these nodes the outcome of measurement. This leads to setting-up of end-to-end entanglement between two nodes A and B that are separated by more than one hop.

Note that conceptually this is similar to link level entanglement. The difference lies in the fact that in multi-hop entanglement set-up, it is performed between two memories, whereas, in link-level entanglement, it is between memory-photon-memory. This has significant difference in performance, since entanglement swapping on photons is limited by 50%, whereas entanglement swapping between memories can be performed deterministically.

This procedure of entanglement swapping between nodes multi-hops apart can be thought of as teleportation between them. In this case, the state of qubit $A'$ is teleported to node B, setting up entanglement between nodes A and B.

### C. Detection of Errors using Purification

As we have explained in Section IV, it is nearly impossible to get a perfectly pure state. In reality, there are imperfections due to system errors and noise. As a result, for a maximally entangled state $|\phi^+\rangle$ shared between two nodes the density matrix $\rho$ can be expressed as:

$$\rho = F|\phi^+\rangle\langle\phi^+| + (1-F)|N\rangle\langle N| \quad (38)$$

Here $F$ represents Fidelity as defined in Eqn. (17) and $N$ denotes noisy quantum state. One simple example of this scenario would be a flip channel where the density matrix in Eqn. (39) above becomes:

$$\rho = F|\phi^+\rangle\langle\phi^+| + (1-F)X|\phi^+\rangle\langle\phi^+|X \quad (39)$$

This can be written as:

$$\rho = F|\phi^+\rangle\langle\phi^+| + (1-F)|\psi^+\rangle\langle\psi^+|$$

The challenge here is to detect such errors. The detection mechanism has to be such that it does not destroy the original entangled state $|\phi^+\rangle$ which is shared between two nodes that are far apart geographically. One of the simplest approaches of detection of quantum errors is what is known as *Purification* [116] [117] [118] [119] [120]. The purification procedure can be accomplished by the setup shown in Fig. 11. It shows a scenario when two nodes A and B share a noisy entangled state between them. In order to perform purification, A and B share another noisy entangled state between them. Both A and B perform CNOT operation on their respective second qubit and shares the resulting outcome with each other. If the measurement outcomes for both A and B are same, they assume that there is no error. Otherwise, they discard the entangled state.

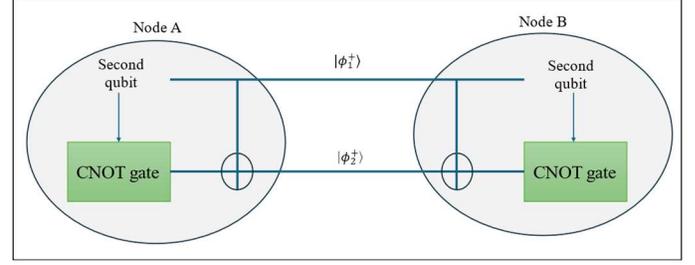

Fig. 11: Detection of errors using purification

This can be understood with an example, where A and B share a pair of maximally entangled states $|\phi_1^+\rangle$ and $|\phi_2^+\rangle$. In the absence of noise, the two-pair system can be expressed as:

$$|\phi_1^+\rangle|\phi_2^+\rangle = \frac{1}{2}(|00\rangle + |11\rangle)(|00\rangle + |11\rangle) \quad (40)$$
$$= \frac{1}{2}(|00\rangle|00\rangle + |11\rangle|00\rangle + |00\rangle|11\rangle + |11\rangle|11\rangle)$$

Upon performing CNOT operation by A, the above expression becomes

$$CNOT_A(|\phi_1^+\rangle|\phi_2^+\rangle) = \frac{1}{2}(|00\rangle|00\rangle + |11\rangle|10\rangle + |00\rangle|11\rangle + |11\rangle|01\rangle) \quad (41)$$

When B performs a CNOT operation, the above expression can be written as:

$$CNOT_{AB}(|\phi_1^+\rangle|\phi_2^+\rangle) = \frac{1}{2}(|00\rangle|00\rangle + |11\rangle|11\rangle + |00\rangle|11\rangle + |11\rangle|00\rangle) \quad (42)$$

From the above expression we can see that, in the absence of noise, both A and B will obtain correlated result. The four measurement outcomes can be obtained with equal probability, but the measurement outcomes will be correlated in A and B.

Next, let us consider that one of these pairs of entangled states is flipped due to noisy channel. Then the two-pair system can be expressed as:

$$|\phi_1^+\rangle|\psi_2^+\rangle = \frac{1}{2}(|00\rangle + |11\rangle)(|01\rangle + |10\rangle) \quad (43)$$
$$= \frac{1}{2}(|00\rangle|01\rangle + |11\rangle|01\rangle + |00\rangle|10\rangle + |11\rangle|10\rangle)$$
$$\Rightarrow CNOT_A(|\phi_1^+\rangle|\psi_2^+\rangle)$$
$$= \frac{1}{2}(|00\rangle|01\rangle + |11\rangle|11\rangle + |00\rangle|10\rangle + |11\rangle|00\rangle)$$
$$\Rightarrow CNOT_{AB}(|\phi_1^+\rangle|\psi_2^+\rangle) = \frac{1}{2}(|00\rangle|01\rangle + |11\rangle|10\rangle + |00\rangle|10\rangle + |11\rangle|01\rangle) \quad (44)$$

TABLE I: PURIFICATION OUTCOMES ON QUANTUM STATES WITH NOISE

| Pair 1 | Pair 2 | Probability | Measurement Outcome | Action |
|---|---|---|---|---|
| $|\phi^+\rangle$ | $|\phi^+\rangle$ | $F^2$ | 00/11 | Keep $|\phi^+\rangle$ |
| $|\psi^+\rangle$ | $|\psi^+\rangle$ | $(1-F)^2$ | 00/11 | Keep $|\psi^+\rangle$ |
| $|\phi^+\rangle$ | $|\psi^+\rangle$ | $F(1-F)$ | 01/10 | Ignore |
| $|\psi^+\rangle$ | $|\phi^+\rangle$ | $F(1-F)$ | 01/10 | Ignore |

Here the measurement outcome on A and B are not correlated. When A measures $|00\rangle$, B will have a different measurement outcome of $|01\rangle$. Thus, the noise or channel errors can be detected by A and B. Table I enumerates the different scenarios how the entangled states $|\phi_1^+\rangle$ and $|\phi_2^+\rangle$ can be affected by noise. From the table, it can be concluded that the purification success probability of detecting errors in entanglement is given by:
$$P_S = F^2 + (1-F)^2$$
Thus, the output Fidelity can be expressed as:
$$F' = \frac{F^2}{F^2 + (1-F)^2} \quad (45)$$
The variation of output fidelity $F'$ with the original fidelity $F$ is shown in the plot depicted in Fig. 12. From the figure, it can be observed that $F' > F$, when $F > 0.5$.

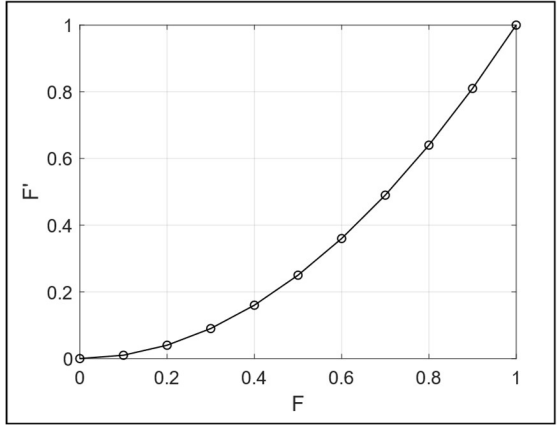

Fig. 12: Variation of output fidelity with the original fidelity

Multiple protocols for entanglement purification have been proposed in literature. The initial protocols [121] [122] used CNOT gates for implementing purification as discussed above. However, implementing CNOT gate in hardware was challenging, and as a solution to this, purification protocols [123] [124] [125] using linear optical elements and nondestructive quantum non-demolition were developed. The papers in [125] [126] [127] use hyperentanglement in order to accomplish entanglement purification.

In general, entanglement purification protocols can be broadly classified into symmetric classification [128], entanglement pumping [129] and greedy purification [130]. Symmetric purification are algorithms that involve entangled pairs with the same fidelity. On the other hand, entanglement pumping [131] is the phenomenon where entanglement between two or more quantum systems is systematically increased or "pumped" over time using repeated interactions or operations. This process involves manipulating the quantum states of the systems in such a way that their entanglement grows stronger with each iteration. Greedy entanglement purification is a method used in quantum information processing to enhance the quality of entangled states by iteratively selecting and purifying the most promising pairs of entangled qubits. An entanglement purification protocol called banded purification scheme is proposed in [132], which is accomplished by selectively purifying entangled qubit pairs within a certain band of entanglement properties. This is reported to be effective when the fidelity of coupled qubits is low, improving the prospects for experimental realization of such systems. The purification protocols above are probabilistic in nature and hence the success of entanglement purification cannot be guaranteed. In scenarios and applications requiring entanglement purification deterministically, the purification schemes proposed in [133] [134] [135] can be utilized. In [133], a purification protocol is proposed that use spatial entanglement and frequency entanglement for correcting bit-flip and phase-flip errors respectively, whereas in the schemes developed in [134] [135], both these kinds of errors are corrected by spatial entanglement. These mechanisms are reported to achieve entanglement purification deterministically. There are works that use machine learning [136] and variational quantum algorithms [137] [138] for heralded purification.

Similarly, several multi-partite entanglement purification protocols have been proposed in recent literature [139] [140] [141]. Two deterministic purification protocols involving multipartite entanglement have been developed in [141] that use spatial and frequency entanglement.

There have been real world implementations and laboratory demonstrations of several purification protocols. The papers [142] [143] [144] [145] [146] report successful execution of entanglement purification using various schemes, viz., spontaneous parametric down conversion, polarization and energy-time hyperentanglement etc. Implementation of deterministic purification protocols are reported in [147] [148].

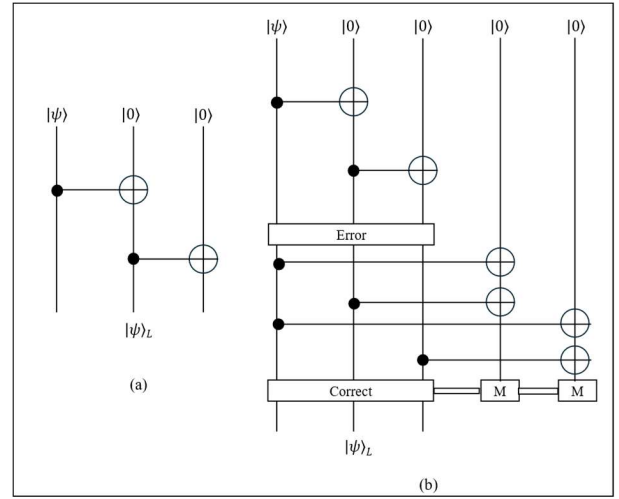

Fig. 13: Quantum Error Correction

### D. Quantum Error Correction

While purification deals with enhancement of quantum entangled states, the goal of Quantum Error Correction (QEC) [149] [150] is to detect and correct errors that occur during quantum computations or quantum communication. It involves encoding quantum information into quantum error-correcting codes (QECCs) [151] [152] [153] [154] to protect against errors caused by noise, decoherence, and other sources of interference. Quantum errors can occur in multiple different forms. For example, coherent errors result from unintended unitary operation on the quantum state, and, these do not necessarily reduce the quantum information. On the other hand, decoherence errors occur when the quantum system interacts with its environment, causing the quantum information encoded in the system to become mixed or entangled with the environment's degrees of freedom. Other kinds of quantum errors include qubit leakage [155] [156] [157] and measurement errors.

Although the basic principles for quantum error correction are the same as in classical error correction schemes, there are certain challenges that need to be addressed for implementing QEC as follows.

1. Unlike classical counterpart, quantum errors cannot be resolved by storing a duplicate copy of the quantum states or qubits. This is because of the '*No cloning Theorem*'.

2. Quantum errors include both bits-flip errors and phase errors. To generalize, quantum errors are continuous in nature, that is, there are errors that can result in shift of quantum bits by an angle.

One of the simplest QEC codes is the 3-qubit code. Though there exists some limitations, this is a good point to understand how QEC can be implemented. The 3-qubit code encodes a single qubit state to a 3-qubit logical state, in order to, detect and correct 1-qubit logical error, as follows.

$$|0\rangle_L \equiv |000\rangle$$
$$|1\rangle_L \equiv |111\rangle$$

Any arbitrary quantum state, thus can be represented as:

$$|\psi\rangle_L = \alpha|0\rangle_L + \beta|1\rangle_L$$
$$\equiv \alpha|000\rangle + \beta|111\rangle \quad (46)$$

The logical bit-flip errors can be detected by computing the distance of the erroneous states from these two codeword states. The number of errors that can be corrected in a quantum state by this codeword is given by:

$$e = \left\lfloor \frac{d-1}{2} \right\rfloor = 1 \quad (47)$$

The schematic circuit to generate such a 3-qubit quantum state from a 1-qubit logical state is shown in Fig. 13 (a). The set-up for physical correction of errors in this setting is depicted in Fig. 13 (b). As can be seen in the figure, error correction is physically done by measurement of two ancillary bits that are 'CNOT-ed' with quantum state whose errors are being detected. Based on the measurement outcome, the location of errors in the multi-qubit state can be detected and corrected by operating it with a Pauli-X matrix. This is demonstrated in Table. II.

TABLE II: ERROR LOCALIZATION

| Error Location | Final State |
|---|---|
| No error | $\alpha|000\rangle|00\rangle + \beta|111\rangle|00\rangle$ |
| Qubit 1 | $\alpha|100\rangle|11\rangle + \beta|011\rangle|11\rangle$ |
| Qubit 2 | $\alpha|010\rangle|10\rangle + \beta|101\rangle|10\rangle$ |
| Qubit 3 | $\alpha|001\rangle|01\rangle + \beta|110\rangle|01\rangle$ |

The above concept is extended in paper [158] to develop a 9-qubit QEC code. In this approach, each logical qubit is encoded to a 9-qubit superimposed state as:

$$|0\rangle_L \equiv \frac{1}{2\sqrt{2}}(|000\rangle + |111\rangle)(|000\rangle + |111\rangle)(|000\rangle + |111\rangle)$$

$$|1\rangle_L \equiv \frac{1}{2\sqrt{2}}(|000\rangle - |111\rangle)(|000\rangle - |111\rangle)(|000\rangle - |111\rangle)$$

In situation where one qubit is decohered, if the measurement is done in the Bell basis, then by taking the majority voting of the three triplets, any error can be deduced and corrected. Note that the Bell basis here is given by $(|000\rangle \pm |111\rangle, |001\rangle \pm |100\rangle, |010\rangle \pm |101\rangle, |100\rangle \pm |011\rangle)$.

The process of decoherence and its detection using the 9-qubit QEC code can be explained as follows. Decoherence of state $\frac{1}{\sqrt{2}}(|000\rangle \pm |111\rangle)$ leads to the following superposed state:

$$\frac{1}{\sqrt{2}}((|a_o\rangle|0\rangle \pm |a_1\rangle|1\rangle)|00\rangle + (|a_2\rangle|0\rangle \pm |a_3\rangle|1\rangle)|11\rangle$$

This can be rewritten in Bell basis as:

$$\frac{1}{2\sqrt{2}}((|a_o\rangle \pm |a_3\rangle)(|000\rangle \pm |111\rangle) + (|a_o\rangle \mp |a_3\rangle)(|000\rangle \mp |111\rangle) + (|a_1\rangle \pm |a_2\rangle)(|100\rangle \pm |011\rangle) + (|a_1\rangle \mp |a_2\rangle)(|100\rangle \mp |011\rangle)) \quad (48)$$

The error correction steps are as follows. The first step is to compare all the three triplets in Bell basis. If these triples are the same, this means that there is no decoherence and nothing needs to be done. If these triplets are not the same, it means that decoherence as occurred and we need restore the encoded qubits to their original state. For example, in Eqn. (48), the output corresponding to the term "$(|a_o\rangle \mp |a_3\rangle)(|000\rangle \mp |111\rangle)$" would represent a wrong sign. Similarly, the output corresponding to "$(|a_1\rangle \pm |a_2\rangle)(|100\rangle \pm |011\rangle)$" would mean that the sign of the qubits is correct but the first qubit is wrong. These outputs are expressed by some quantum state of the ancilla, which then is measured. Note that error correction has evolved into a unitary transformation since we're not solely rectifying the error but also "measuring" it.

Multiple variants of this code have been developed in the literature [159] [160]. For example, the 5-qubit code proposed in [159] and 4-qubit code in [160] are sophisticated versions of the Quantum Error Correction Code. A major development in this area came from the development of surface code [161] [162] [163] which was inspired from *Toric* code proposed in [164] [165]. Surface code is particularly well-suited for quantum computing architectures where qubits are arranged in a two-dimensional lattice, such as superconducting qubits or topological qubits. The surface code offers high error detection and correction capabilities while requiring relatively simple operations. The working of surface code can be summarized in the following few steps.

Qubit Arrangement: Qubits are first arranged on a two-dimensional lattice, typically in a square or rectangular grid. Each qubit is surrounded by four neighboring qubits, forming a "plaquette" or "vertex" configuration.

Encoding: Logical qubits are encoded into multiple physical qubits in such a way that errors can be detected and corrected. The surface code typically encodes one logical qubit into multiple physical qubits, often using a 2D arrangement of qubits.

Measurement Syndrome: Quantum error correction in the surface code relies on the measurement of stabilizer operators, which are products of Pauli operators (X, Y, Z) that commute with the encoded logical operators. By measuring stabilizer operators associated with sets of qubits (e.g., plaquettes or vertices), a syndrome is obtained that indicates the presence and location of errors.

<u>Error Detection</u>: Errors are detected by measuring the stabilizer operators associated with each plaquette or vertex. Any deviation from the expected stabilizer measurement outcomes indicates the presence of errors.

<u>Syndrome Extraction</u>: The syndrome is extracted from the stabilizer measurement outcomes, revealing the type and location of errors in the encoded qubits.

<u>Error Correction</u>: Based on the syndrome, error correction operations are applied to rectify the errors. Correction operations are typically applied by performing controlled-X or controlled-Z gates between qubits based on the syndrome information.

Different versions of sophisticated error correction protocols have been developed in recent literature. One of the much explored techniques for developing logics for quantum error correction is the usage of machine learning frameworks [166] [167] [168] [169] [170].

### E. Routing in Quantum Networks

Till this point in the discussion on long distance communication, we have seen how link level entanglement can be established and how that can be leveraged to perform entanglement swapping between two nodes that are multiple hops apart. We also observed how to detect errors and noise in scenarios of shared entangled state. One important point that needs to be addressed here is how to find the route between the source to destination, in the presence of multiple hops between them.

The concept of routing in quantum networks is the same as classical networks. The only difference here is the cost definition of a route. In traditional routing approaches, the cost is often defined by the path length, channel reliability etc. In quantum network, one additional cost needs to be considered, which is the cost of entanglement swapping. We have already explained the loss of photons in fibers and how this can be handled by sending multiple copies of photons. In other words, there is a time required for setting up an entanglement between nodes. Thus, one way of defining cost is the time required for setting up a Bell pair for a given threshold fidelity. Based on this cost, modified versions of the existing routing algorithms such as Bellman Ford or Dijkstra's routing algorithm can be used to route the packets from source to destination.

The most widely used quantum routing protocols [171] [172] [173] [174] [175] [176] [177] [178] [179] [180] rely on network time-slotting and time synchronization. During each time-slot, the entanglement generation and entanglement swapping take place and it is assumed that entangled states are perfect within a time slot. The optimal route for the quantum routing algorithms is defined in [181] as the path that has the largest number of end-to-end entangled states in a time slot, the highest fidelity, the lowest consumption of entangled resources, and the shortest delay. Note that, for a quantum network with mesh topology, a single route that satisfies all these conditions is difficult to attain. Hence, multiple application-specific target routes are obtained to satisfy the performance metric.

As an example, probabilities of entanglement generation and swapping can be used for defining throughput ($S$) of a route, as is defined in [182]:

$$S = q^h \sum_{i=1}^{W} i \cdot (P_h^i) \qquad (49)$$

$$P_h^i = P_{h-1}^i \sum_{l=i}^{W} \binom{W}{l} p_h^l (1-p_h)^{W-l}$$
$$+ (\binom{W}{i} p_h^i (1-p_h)^{W-i}) \sum_{l=i+1}^{W} P_{h-1}^l$$

Here, $P_h^i$ is the probability of each of the first $h$ hops of the path having $i$ successful entangled states; $p_h$ is the success probability of HEG on the $h^{th}$ hop; $q$ is the success probability of Entanglement Swapping; $h$ is the number of hops and $W$ is the number of entangled states used by each link.

Similarly, in the Q-Leap algorithm developed in [180], the authors use fidelity of a path to decide next hop. The end-to-end fidelity of a route with $h$ number of hops is defined as:

$$F = \prod_{i=1}^{h} F_i \qquad (50)$$

If the route costs (or inversely, the throughput) are defined using the metrics above (Eqn. 50), then the cost of the entire route is different from the sum of the costs of individual route. In other words, the route cost is not additive. Thus, the raw version of Dijkstra's routing algorithm cannot be used directly. As a result, the extended version of Dijkstra's algorithm [173] is used in this case.

The Extended Dijkstra's Algorithm (EDA) does not require the route cost to be additive but needs to be monotonic. A routing metric function (equivalent to routing cost) $e$ is used to find the best possible route. EDA gives the path with the maximum value of the evaluation metric between the source and destination.

In a manner akin to the classical Dijkstra algorithm, the Extended Dijkstra Algorithm (EDA) similarly establishes an optimal spanning tree originating from the source node, denoted as $S$. Initially, only $S$ is included in the set of visited nodes. The evaluation score from $S$ to an unvisited node $X$ is initialized to 0 or to the evaluation score $e(S,X)$ of the edge $(S,X)$ if $S$ and $X$ are neighboring nodes. During each iteration, the node $Y$ with the highest evaluation score with respect to $S$ is appended to the set of visited nodes, and the evaluation scores from 's' to any other node $X$ are adjusted if $X$ and $Y$ are neighbors.

Note that, in contrast to the classical Dijkstra algorithm, when updating the path by adding one hop, it might necessitate re-evaluating the entire path instead of merely incorporating the cost of a link. To circumvent the costly recalculation for path updates, different optimization techniques can be implemented as is shown in [182].

One notable point in the EDA algorithm is the ignoring of unreasonably longer routes. An upper-bound threshold $h_m$ for the hop count of routes is set to ensure bounded searching in EDA. Moreover, in EDA, in addition to the major paths found using the concept described above, the remaining qubits and channels are utilized to construct recovery paths, each of which ends at two nodes (denoted as switch nodes) on a single major path. The switch nodes should be no more than $k$ hops away on a major path, where k is the link state range.

Another alternative to avoid such complexities used in Extended Dijkstra's Algorithm, some researchers used other metrics for routing. As an example, in [183], the authors use the inverse of the throughput of a single link after purification as a routing metric. This measure is additive, and as such, Dijkstra's routing algorithm can be directly used.

## IX. QUANTUM INTERNET

The idea of the quantum internet [184] [185] [186] [187] [188] is to enable communication between two quantum computers located at any point on the earth [189] (Fig. 14). The current developments of quantum communication would require its coexistence with classical communication to realize a full-potential quantum internet. However, there are research works [190] that envision quantum internet as a standalone system relying only on quantum channel connecting the edge quantum nodes. Nevertheless, for the latter to be functional, we need to understand and resolve the challenges of the first.

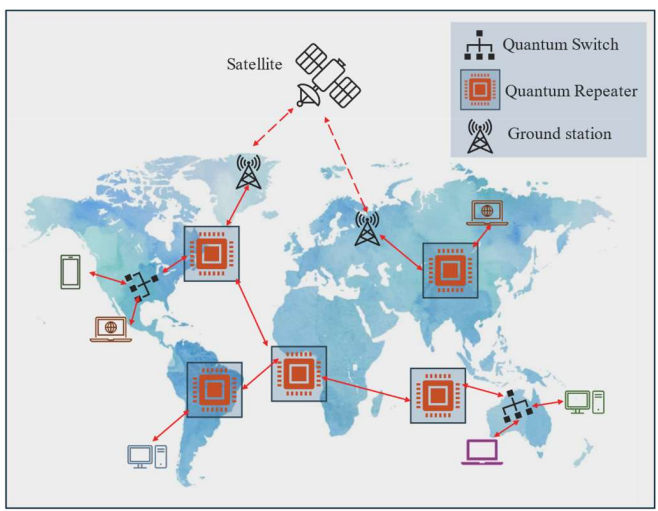

Fig. 14: A schematic representation of the Quantum Internet

We have discussed all the components that are necessary for real-world realization quantum internet. Here, let us put all these components together and understand what challenges that we need to overcome for successful implementation of the large-scale quantum internet.

### A. Components of Quantum Internet

The first and foremost component that is indispensable for quantum internet is a physical channel that supports transmission of qubits. One such example is an optical fiber that is used in classical communication today. The next necessary component is a way or mechanism to enable long distance quantum communication. We have already discussed the approaches to achieve this. One main element of enabling long distance quantum communication is the use of quantum repeaters. Such repeaters are needed to be placed at certain intervals over the optical fiber connection, to allow the qubits to be transmitted over a long distance. Last, but not the least, we need end devices, which are, in general, quantum processors that can generate and measure qubits and quantum states.

<u>Physical Channel</u>: The most prevalent form of communication channel for transmitting qubits is optical link [191], [192], [193], [194]. However, the experimental validation of satellite quantum channel is demonstrated in [195] [196]. As pointed out in [197], the future quantum internet may use both these forms of channels depending on the application requirements. Whatever the kind of channels are, these should have minimal photon loss and should exhibit low decoherence. Although there are photon heralding protocols that are used to ameliorate photon loss, photon loss would have adverse effect on quantum communication rates. There are entanglement distillation protocols [198], [199] that are used to overcome the effect of decoherence.

<u>End Devices</u>: The first major requirement for the end nodes is to reliably store quantum state information when the entanglement is set up. Several mechanisms of photon-mediated point-to-point entanglement have been proposed in literature. Some of these approaches include generation of well-controlled photons using semiconductor quantum dots [200] [201], trapped ions [202], nitrogen-vacancy centers in diamond [203]. The paper [204] demonstrates the creation of high-fidelity photonic quantum states over a distance of 1.3 km. The major challenge of extending these point-to-point entangled links to an actual quantum network, as pointed out in [197], is the reliable storage of the quantum states in the nodes. Although, the quantum states created by the aforementioned techniques have a very high coherence time, however, there are scenarios where unwanted couplings can significantly impact the coherence of a memory qubit during operations on another qubit within the same node [205], [206]. One solution used to overcome this involves utilizing different types of qubits within a node. For example, trapping various ion species allows for individual addressing based on their distinct electronic transition frequencies [207], [208], [209]. Similarly, carbon-13 nuclear spins near a diamond N-V center offer a robust register of memory qubits that do not interact with the laser control fields on the N-V electron spin [210]. Recent experiments have demonstrated that such hybrid network nodes enable the generation of remote entangled states, facilitating entanglement distillation [211]. Integrating multiple robust memories into a multiqubit network node could bring us closer to achieving the highest stages of the quantum internet. However, a key challenge lies in realizing a robust and highly efficient converter with a high signal-to-noise ratio. Alternatively, end nodes could consist of a quantum processor with qubit frequencies in the microwave domain, such as a superconducting qubit circuit, paired with a microwave-to-optical conversion process.

<u>Quantum Repeaters</u>: Similar robustness in storing and processing quantum states is required in quantum repeaters as needed by the end nodes. But this requirement is significantly less drastic as compared to the end nodes. This is because the time to store the qubits reliably is limited to the time of establishing entanglement between the one-hop nodes of the repeaters. As for example, an on-demand quantum memory can be developed by an ensemble of atoms and ions in gas or solid phase, as shown in [212]. Storage and on-demand retrieval of a photon's quantum state with low efficiency are shown in [213], [214], [215], [216] using heralded quantum memory. Research is going on to increase the coherence time of quantum states and increase the efficiency of on-demand state retrieval. A different technique used in designing quantum repeaters is by encoding multiple photons with the same quantum states. The idea here is that the repeaters can retrieve the original quantum state by means of quantum error correction and ameliorate the effects of photon loss and decoherence [217], [218], [219], [220]. This technique achieves a high communication rate, however, one limitation here is that this mechanism requires deterministic generation of multi-photons quantum state, which is far from implementation till now [221]. The work in [222] proposed a framework to circumvent the inherent limitations of quantum repeaters by integrating classical repeaters into quantum internet. Security is taken care of by incorporating quantum secure direct communication (QSDC) principle [223] [224] into the classical repeaters. In the proposed model, the encrypted data generated from a quantum-resistant algorithm is propagated through QSDC across network nodes. Each node reads the data and subsequently relays it to the next node in the sequence. At the classical repeaters, the information is repotted to remain protected by the proposed quantum-resistant algorithm. The proposed architecture is shown to work in an end-to-end quantum communication link, and it is reported to be able to detect and prevent eavesdropping for the next-generation internet.

*B. Recent Trends in Quantum Internet*

Given the limitations and challenges pointed out above, quantum internet is still at an infant stage of development. There are multiple research that aim at overcoming these challenges both from the perspective of hardware architecture and network protocols. The paper [225] has proposed a quantum internet architecture that relies on the Quantum Recursive Network Architecture (QRNA), using RuleSet-based connections established by a two-pass connection setup. This paper proposes an end-to-end architecture for establishing entanglement using minimum hardware requirements. The proposed scalable architecture is reported to achieve multi-partite entanglement and would assist quantum error correction.

The papers [181], [226] [227] [228] [229] [230] [231] [232] [233] [234] [235] [236] present the layered architecture of the protocol stack for quantum internet. In the paper [181], the quantum physical layer is defined as the one that furnishes the quantum internet with diverse quantum assets, encompassing entanglement generation within a single hop, emission of single photons, quantum memory, interfaces between quantum light and matter, and conversion of quantum frequencies. Although there are physical layer protocols [237] [238] [239] [240] [241] [242] [243] for achieving these, as we have discussed in Section IX A, the main hurdles for real-world deployment of the system are as follows.

1. Quantum Interface Problems: These include problems such as information transfer and entanglement between matter qubits on nodes and photons as depicted in [244] [245] [246] [247] [248] [249] [250] [251] [252] [253] [254].

2. The quantum devices available currently are not sophisticated enough to generate qubits with very high generation rate and to maintain their quantum properties long enough [255] [256] [257] [258] [259] [260] [261] [262] [263] [264] [265] [266]. The present fidelity and storage time of the qubits need significant improvement in order for the system to be made large-scale deployment worthy.

The quantum link layer, as defined in paper [181], aims at establishing and maintaining a robust entanglement between two nodes. The primary functionality of the link layer is to notify failure if the link conditions are not suitable for establishing entanglement. This is made sure by observing fidelity values and memory status of the nodes. The link layer also executes the different Quantum Error Correction (QEC) and Entanglement purification protocols as we discussed in the previous sections. In Section VIII, we have given an overview of the existing QEC protocols. In recent years, studies have showcased the practical implementation of diverse QEC codes across a range of hardware platforms, including nuclear magnetic resonance (NMR) [267], [268], trapped ions [269] [270] [271] [272] [273] [274] [275] [276] [277] [278], diamond [279] [280] [281], silicon [282], superconducting circuits [283] [284] [285] [286] [287] [288] [289] [290] [291] [292] [293], and photons [294] [295]. The ongoing research in this field is in developing QEC protocols for heterogeneous networks, as shown in [296], where a scheme of heterogeneously encoding Bell pairs was devised. In that scheme, two halves of the Bell pair were encoded using Steane code [297] and surface code respectively. Similarly, current research on entanglement purification protocols is to make them robust to decoherence, noise and other errors. In [298] [299], the authors develop entanglement access control (EAC) to solve the contention problem arising in accessing a multipartite entangled resource.

Coming into the routing layer, as we have seen in Section VIII, the different routing protocols for quantum communication use various performance metrics into consideration, such as, throughput, latency, distance, number of repeaters, entanglement cost, fidelity, node qubit capacity. Recent literature has explored advanced quantum functionalities that leverage quantum control registers to enable network devices to manipulate coherent superpositions of various tasks. Such capabilities facilitate the preparation of superpositions of different target state configurations among network nodes and enable quantum information transmission across superposed trajectories or to multiple destinations. The concept of quantum addressing has emerged as a promising approach to achieve full quantum functionality within quantum networks. Recent experiments have demonstrated the inherent robustness of quantum communication using superposed trajectories, underscoring the potential viability of such approaches. The current routing challenges lie in finding improved quantum network models and appropriate routing metrics, designing better routing protocols, handling resource competition, and developing improved methods for calculating waiting times.

Researchers have also developed transport layer protocols for quantum communications [300] [301]. These mainly deal with three broad issues: quantum congestion control, retransmission and quantum channel quality detection. These transport layer functionalities and the relevant protocols have already been detailed in Section VIII. The main open problems and challenges for developing fully-functional protocols in this layer for quantum internet include:

1. The current transport layer protocols discard entangled states and quantum data under congestion. This is not useful from two perspectives. First, given the current limitations of the lower layers, setting up end-to-end entanglement is a challenge. Second, discarding entanglement and quantum information may lead to precious data loss, as quantum states cannot be recreated, because of the property of no-cloning theorem. Therefore, research should be done to find alternatives to discarding the entangled quantum states in the transport layer.

2. The second open problem in this layer lies in exploring advanced quantum retransmission protocols to optimize reliability and efficiency. Existing protocols driven by quantum secret sharing require two successful transmissions to complete data transfer, even in scenarios without any packet loss [181]. This leads to large transmission time which needs to be resolved.

3. Research needs to be done in developing strategies to minimize packet loss in quantum networks. Considering the current stage of development of quantum communication, packet loss is much more detrimental here as compared to classical communication. Quantum states cannot be recreated and also these require processors with very high computational complexity.

Finally, thought and effort must be put forward for developing the application layer protocols, which is at its infancy at present. As reported in paper [181], the present state of quantum applications is limited, given the limitations in generations and processing capabilities of the hardware. In paper [302], the authors estimate that factoring a 2048-bit RSA integer would require 20 million physical qubits. On the other hand, the most advanced quantum computer at present has the capability of handling a maximum of 1121 qubits. Thus, the development of the application layer protocols would be heavily guided by the lower layer hardware and software support. However, it is time to find real world applications, where the quantum properties can be exploited to solve them, while keeping these within the reach of moder capacity of quantum computers.

Recent research has also investigated security and different forms of attacks on the quantum internet [303] [304] [305] [306]. The study developed in [303] has focused on the attacks on quantum repeaters mainly from the perspective of confidentiality, integrity, and availability. Similarly, the work in [305] has proposed a lightweight defense mechanism for quantum attacks in an IoT network. The following key aspects on security can be commented from perusing the literature. Quantum repeater systems offer significant advantages in terms of confidentiality due to their ability to detect eavesdroppers, allowing for breach detection. Quantum tomography [307] [308], while sacrificing some Bell pairs for network monitoring and fidelity optimization, can also aid in eavesdropper detection when the sacrificed portion is chosen randomly. However, integrity and availability concerns in quantum repeater systems are akin to those in classical networks, with threats potentially targeting the classical computing hardware within repeaters. To ensure security, attention must also be given to securing the classical components of the system, such as the quantum node's classical parts and network services, which are susceptible to mixed attacks combining quantum and classical elements.

Finally, there are certain network design challenges that are of interest to quantum communication engineers. The work in [309] investigate the scalability of quantum internet for scenarios when the users are separated over long distances or within a small space over a processor or circuit within a same framework. Similarly, the authors in [310] consider computational complexity for entanglement distributions in quantum networks. In [311], the authors provide a set of guidelines on using entanglement distribution based on user requirements and hardware configuration.

## X. SUMMARY: RESEARCH SCOPE, CHALLENGES AND SIMULATION TOOLS

This paper delves into the core fundamentals of quantum communication and provides a comprehensive review of the field relevant to the state-of-the-art communication technologies. In order to provide the readers with a better understanding of the recent progress of the topic, we start with the first principles covering the details on quantum mechanics and quantum computing. The key similarities and differences between classical and quantum communication are pointed out and it is explained how these two counterparts coexist. It should be emphasized here that quantum communication should not be visualized as a replacement for classical communication. Rather, quantum communication is a layer embedded on top of the classical counterpart with its own set of rules, logic and protocols.

The paper discusses the fundamental concepts specific to quantum communication, including superposition, entanglement and teleportation. It demonstrates how these concepts can be leveraged for achieving secure and reliable quantum communication. Using the unique quantum properties, the several benefits relevant to modern day communication can be accomplished. Two such noteworthy benefits include enhanced security and improved information capacity. Quantum communication protocols leverage the fundamental properties of qubits, such as superposition and entanglement, to ensure secure transmission of information. Similarly, quantum systems can encode and process information using multiple degrees of freedom, leading to a higher information capacity compared to classical systems.

This paper also covers the implementation advancements for execution of the quantum communication protocols. It is explained how all the basic operations of quantum computing can be achieved from hardware implementation perspective. The paper provides circuit-level details required for execution of link-level entanglement and successful realization of teleportation and long-distance quantum communication. Handling of quantum errors using purification and quantum error correction is demonstrated.

Finally, we cover the current-day progress of quantum internet, particularly focusing from the perspective of network layer architecture. The unique attributes of quantum mechanics have been leveraged for quantum computing and communication in variety of fields and domains, such as, vehicular communications [312] [313], biomedical applications [314], social networks [315], electric grids [316], etc. The paper also summarizes the recent trends in this domain with a multitude of research endeavors aimed at addressing various open problems and challenges in the field. These challenges span from hardware architecture to network protocols, with a focus on achieving scalable and efficient quantum communication systems. For instance, proposed architectures like QRNA aim to establish end-to-end entanglement using minimal hardware while ensuring multi-partite entanglement and supporting quantum error correction. However, key challenges persist, such as quantum interface problems that hinder efficient information transfer and entanglement between matter qubits and photons. Additionally, the limitations of current quantum devices, including low qubit generation rates and short quantum property maintenance times, necessitate significant improvements for large-scale deployment. The quantum link layer plays a crucial role in maintaining robust entanglement between nodes, yet challenges remain in implementing effective QEC protocols across heterogeneous networks and developing entanglement purification protocols resilient to decoherence and other errors. Furthermore, advancements in the routing layer are vital, requiring improved network models, routing metrics, protocols, and resource management strategies to optimize performance and reliability. The transport layer, focused on congestion control, retransmission, and channel quality detection, faces challenges in discarding entangled states under congestion, optimizing retransmission protocols for efficiency, and minimizing packet loss in quantum networks. Finally, there are open research challenges for developing application layer protocols that leverage quantum properties while remaining compatible with current quantum computing capacities. This would pave the way for real-world applications that harness the full potential of quantum communication technologies.

There are many simulation tools and software that has helped and would help researchers to contribute to the field of quantum communication. Although there are many advancements that would be incorporated as we progress more in this field, these tools are of tremendous importance for implementing key quantum concepts. Quantum Information Science Kit (Qiskit) is an open-source quantum computing software development framework provided by IBM [317]. While primarily focused on quantum computing, Qiskit also includes tools and libraries for simulating quantum circuits and protocols, making it useful for quantum communication simulations. QuTiP (Quantum Toolbox in Python) [318] is an open-source quantum computing simulation library written in Python. It is useful for execution of simulating functionalities of quantum mechanics, quantum optics, and quantum information science. Another quantum computing software developed by Google is known as Cirq [319] that can be used for creating, simulating, and running quantum circuits on Google's quantum processors. ProjectQ [320] and Quantum++ [321] are two other simulation tools for implementation of quantum algorithms and circuits.